\documentclass[9pt,twocolumn,twoside]{pnas-new}
\templatetype{pnasresearcharticle} % Choose template 
\usepackage{graphicx}
\usepackage{graphics}
\usepackage{amsfonts}
\usepackage{epstopdf}
\usepackage{epsfig}
\usepackage{dcolumn}
\usepackage{bm}
\usepackage{latexsym,amsmath,amssymb,euscript}
\usepackage{color}
\usepackage{amsbsy}
\usepackage{times}
\usepackage{multirow}
\usepackage{subfigure}
\usepackage{bbm}
\def\ra{\rangle}
\def\la{\langle}
\def\up{\uparrow}
\def\dn{\downarrow}
\def\Hc{{\rm H.c.}}
\title{Weyl-Kondo Semimetal in Heavy Fermion Systems}

\author[a,1]{Hsin-Hua Lai}
\author[a,1]{Sarah E. Grefe} 
\author[b]{Silke Paschen}
\author[a]{Qimiao Si}

\affil[a]{Department of Physics and Astronomy \& Rice Center for Quantum Materials, Rice University, Houston, Texas 77005, USA}
\affil[b]{Institute of Solid State Physics, Vienna University of Technology, Wiedner Hauptstra{\ss}e 8-10, 1040 Vienna, Austria}

\leadauthor{Lai} 

\significancestatement{While electronic states with nontrivial topology have traditionally been known in insulators, they have been evidenced in metals during the past two years. Such Weyl semimetals show topological protection while conducting electricity both in the bulk and on the surface. An outstanding question is whether topological protection can happen in metals with strong correlations. Here, we report the first theoretical work on any strongly correlated lattice model to demonstrate the emergence of a Weyl-Kondo semimetal. We identify Weyl fermions in the bulk and Fermi arcs on the surface, both of which are associated with the many-body phenomenon called Kondo effect. We determine a key signature of this Weyl-Kondo semimetal, which is realized in a newly discovered heavy fermion compound.}

\authorcontributions{All authors contributed to the research of the work and the writing of the paper. }
\authordeclaration{The authors declare no competing financial interests.}
\equalauthors{\textsuperscript{1}H.-H. L. and S. E. G. contributed equally to this work.}
\correspondingauthor{\textsuperscript{2}To whom correspondence should be addressed. E-mail: hl53@rice.edu (H.-H. L.); qmsi@rice.edu (Q. S.)}


\begin{abstract}
Insulating states can be topologically nontrivial, a well-established notion that is exemplified by the quantum Hall effect  and topological insulators. By contrast, topological metals have not been experimentally evidenced until recently. 
In systems with strong correlations, they have yet to be identified. Heavy fermion semimetals are a prototype of strongly correlated systems and, given their strong spin-orbit coupling, present a natural setting to make progress. Here we advance a Weyl-Kondo semimetal phase in a periodic Anderson model on a noncentrosymmetric lattice. 
The quasiparticles near the Weyl nodes develop out of the Kondo effect, as do the surface states that feature Fermi arcs. We determine the key signatures of this phase, which are realized in the heavy fermion semimetal Ce$_3$Bi$_4$Pd$_3$. Our findings provide the much-needed theoretical foundation for the experimental search of topological metals with strong correlations, and open up a new avenue for systematic studies of such quantum phases that naturally entangle multiple degrees of freedom.
\end{abstract}

\dates{This manuscript was compiled on \today}
\doi{\url{www.pnas.org/cgi/doi/10.1073/pnas.XXXXXXXXXX}}

\begin{document}

% Optional adjustment to line up main text (after abstract) of first page with line numbers, when using both lineno and twocolumn options.
% You should only change this length when you've finalised the article contents.
\verticaladjustment{-2pt}

\maketitle
\thispagestyle{firststyle}
\ifthenelse{\boolean{shortarticle}}{\ifthenelse{\boolean{singlecolumn}}{\abscontentformatted}{\abscontent}}{}

% If your first paragraph (i.e. with the \dropcap) contains a list environment (quote, quotation, theorem, definition, enumerate, itemize...), the line after the list may have some extra indentation. If this is the case, add \parshape=0 to the end of the list environment.
\dropcap{S}trongly correlated electrons represent a vibrant field that continues to expand its horizon. In heavy fermion systems, strong correlations make their ground states highly tunable and give rise to a rich phase diagram that features antiferromagnetic order, Kondo-screened and other paramagnetic phases, and beyond-Landau quantum phase transitions \cite{ColemanSchofield2005,SiSteglich2010}.
In the simplest cases, these systems can be considered in terms of the local moments originating from the $f$-electrons that Kondo couple to the spins of the conduction electrons. The interaction generates the Kondo spin-singlet ground state; the ensuing entanglement with the conduction electrons converts the local moments into quasiparticles that can hybridize with the conduction electrons. This leads to a metal with a large, strongly renormalized effective carrier mass, which is the hallmark of the heavy fermion system classification. The resulting state could be a heavy fermion metal or a Kondo insulator depending on whether the chemical potential lies within or falls between the hybridized bands \cite{AeppliFisk,SiPaschen2013,Dzero2010}. Electronically intermediate between the two cases are heavy fermion semimetals ~\cite{Guritanu2013,Sunderman2015,Luo2015,Dzsaber2016,Mason1992,Sto16.1,Feng2016,Pixley2017}. Several of these have a broken inversion symmetry, including CeRu$_4$Sn$_6$~\cite{Guritanu2013,Sunderman2015}
and Ce$_3$Bi$_4$Pd$_3$~\cite{Dzsaber2016}.

Semimetal systems are being theoretically studied in the noninteracting  limit with spin-orbit coupling, which plays an essential role in obtaining  topological phases of electronic matter~\cite{Hasan_Kane_RMP, Qi_Zhang_RMP, Moore2010,Bernevig13}. The Weyl semimetal in three dimensions (3D) was recently evidenced experimentally
~\cite{WSM_TaAs1,WSM_TaAs2,WSM_optical}. It possesses  bulk excitations in the form of chiral fermions, with massless relativistic dispersions near pairs of nodal points in the momentum space, as well as surface states in the form of Fermi arcs~\cite{AMV_RMP,Hosur13,Wan2011,Burkov2011}. Because both the bulk and surface states are gapless, one can expect that the Weyl semimetals are particularly susceptible to the influence of electron correlations. Moreover, strong correlations in non-perturbative regimes typically mix different degrees of freedom in generating low-energy physics; thus, in any strongly correlated Weyl semimetal, the low-energy electronic excitations are expected to involve degrees of freedoms such as spin
moments, which may be harnessed for such purposes as information storage and retrieval.
 
In this paper, we report the discovery of a Weyl-Kondo semimetal (WKSM) phase in a concrete microscopic model on a 3D noncentrosymmetric lattice. This model contains the strongly correlated $4f$ electrons and a band of conduction $spd$ electrons, respectively. It is realistic in that it captures the inversion-symmetry breaking and spin-orbital coupling in a tunable way. In the regime where the electron-electron repulsion is much larger than the width of the conduction-electron band, the interaction-induced renormalization factor can be very large. In addition, because the inversion-symmetry breaking term, spin-orbit coupling and other electronic couplings are renormalized in very different ways, it is {\it a priori} unclear whether any Weyl state can be realized in a robust way. Our work advances an affirmative answer in this well-defined
microscopic model. Moreover, we demonstrate the key signatures of the 
WKSM phase, which turn out to be realized in several new heavy fermion compounds.

\begin{figure}[t]
 \centering
    \subfigure[]{\label{Fig:lattice}\includegraphics[width= 1 in]{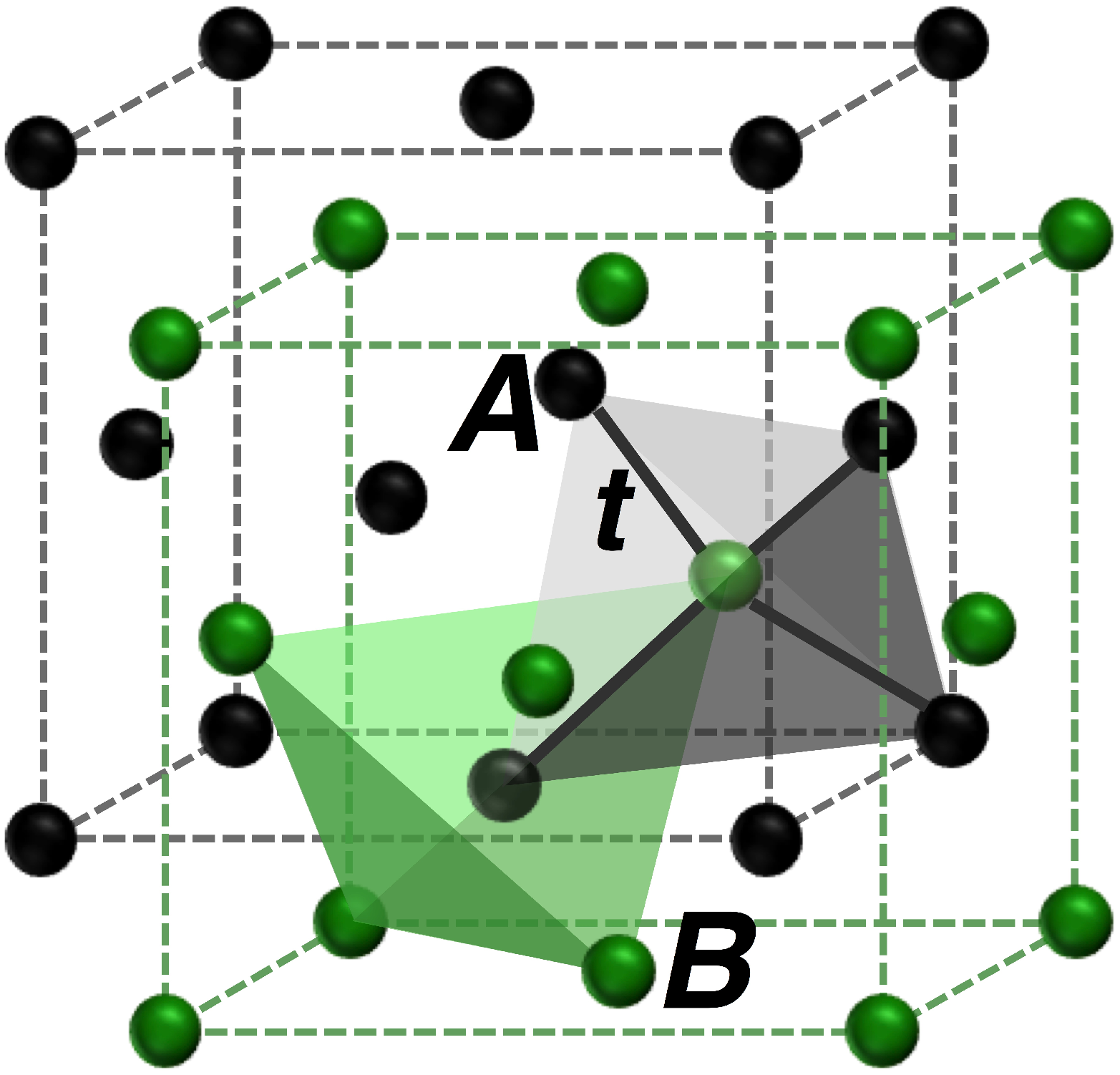}}
  \subfigure[]{\label{Fig:inv_tet}\includegraphics[width= 1 in]{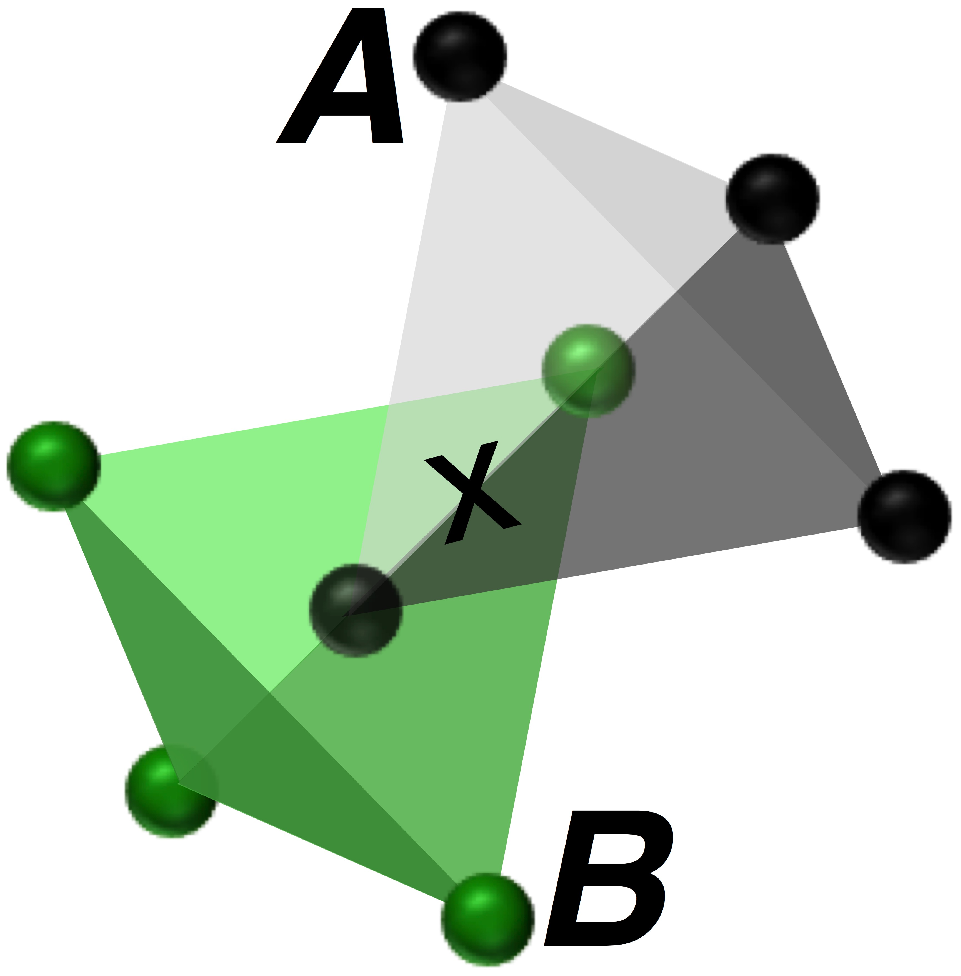}}
  \subfigure[]{\label{Fig:3D_BZ}\includegraphics[width= 1.1 in]{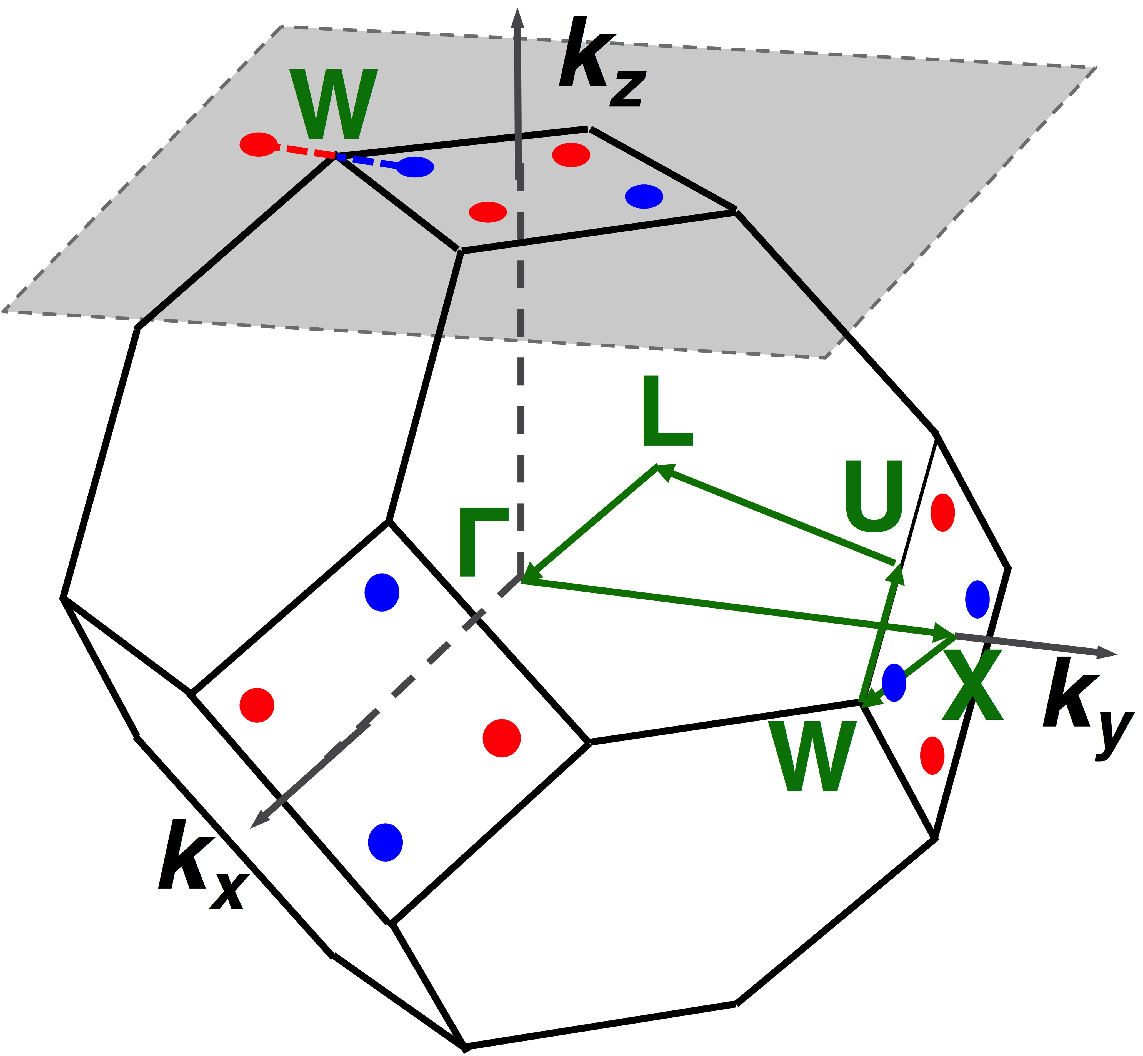}}
    \caption{{\bf The 3D noncentrosymmetric lattice and associated Brillouin zone.}  
		{\bf (a),} Diamond lattice with hopping $t$ and onsite energy $\pm m$ differentiating $A,~B$ sublattices. The solid lines connect nearest neighbors;
		{\bf (b),} Interlocking tetrahedral sublattice cells illustrating how the distinction between the A and B sublattices (Zincblende structure) invalidates the inversion center lying on the point marked ``X'';
		{\bf (c),} The Brillouin zone (BZ) of the diamond lattice, with Weyl nodes shown in blue/red, 
		and high symmetry contour used for Fig.~\ref{Fig:strong_spec}
		in green.}
\label{Fig:lattice_BZ}
\end{figure}

%%%%%%%%%%%%%%%%%%%%%%%%%%%%
The Hamiltonian for the periodic Anderson model to be studied is
\begin{eqnarray}
H = H_d +H_{cd} + H_c  .
 \label{Eq:H_pam}
\end{eqnarray}
For a proof-of-concept demonstration, we consider a cubic system in which the breaking of inversion symmetry can be readily incorporated.
This is a diamond lattice, which comprises two interpenetrating face-centered cubic lattices A and B (Fig.~\ref{Fig:lattice}). We have chosen this lattice because it is nonsymmorphic and, in the case of non-interacting electrons, band touching 
is enforced by its space group symmetry~\cite{Kane_3ddirac}. The model contains $d$ and $c$ electrons, 
corresponding to the physical $4f$ and $spd$ electrons, respectively.
The first term, $H_d$, describes the $d$ electrons with an energy level $E_d$ and a Coulomb repulsion $U$. The coupling between the two species of electrons is described by a bare hybridization of strength $V$. The conduction-electron Hamiltonian $H_c$ realizes a modified Fu-Kane-Mele model~\cite{FKMmodel07}. Each unit cell has four species of conduction electrons, denoted by sublattices $A$ and $B$ and physical spins $\up$ and $\dn$: $\Psi^T_{\bf k}=\begin{pmatrix} c_{{\bf k}\up,A}& c_{{\bf k}\up,B} & c_{{\bf k}\dn,A} & c_{{\bf k}\dn,B}\end{pmatrix}$. There is a nearest-neighbor hopping $t$ (chosen as our energy unit), and a Dresselhaus-type spin-orbit coupling of strength $\lambda$.
The  broken inversion symmetry, Fig.~\ref{Fig:inv_tet}, is captured by an onsite potential $m$ that staggers between the A and B sublattices~\cite{Ojanen13,Bernevig13}. The band basis is arrived at by applying a canonical transformation on
 $H_c$ written in the sublattice and spin basis. It corresponds to a pseudospin basis~\cite{Ojanen13}, defined by the eigenstates  $|\pm D\ra$.
 We fix the $d$-electron level  to below the conduction-electron band, and consider the case of 
 a quarter filling, corresponding to one electron per site.
Further details of the model are given in the Materials and Methods section.

%%%%%%%%%%%%%
%%%%%%%%%%%%%%%%%%%%%%%%%%%%%%%%%%%%%%%

We consider the regime with the onsite interaction $U$ being large compared to the bare $c$-electron bandwidth ($U/t \rightarrow \infty$). 
We approach the prohibition of $d$ fermion double occupancy by an auxiliary-particle method~\cite{Hewson_Kondo_book}: $d^\dagger_{i\sigma} = f^\dagger_{i\sigma} b_{i}$. Here, the $f^\dagger_{i\sigma}$ ($b_i$) are fermionic (bosonic) operators, 
which satisfy a constraint that is enforced by a Lagrange multiplier $\ell$. This approach leads to a set of saddle-point equations, 
where $b_i$ condenses to a value $r$, which yields an effective hybridization between the $f$-quasiparticles 
and the conduction $c$-electrons. The details of the method are described in the Materials and Methods section. 

\begin{figure}[t] 
   \centering
   \includegraphics[width=2 in]{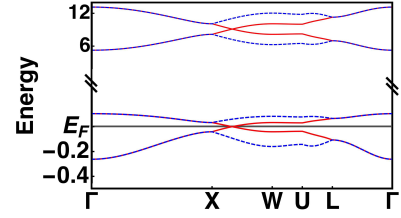} 
   \caption{{\bf Energy dispersion of the bulk electronic states.} Shown here is the energy {\it vs.} wavevector ${\bf k}$ along a high symmetry path in the BZ, defined in Fig.~\ref{Fig:3D_BZ}.
   The bottom four bands near $E_F$ show
   a strong reduction 
   in the bandwidth.,
   The bare parameters are $(t, \lambda, m, E_d, V)=(1,0.5, 1, -6, 6.6)$.
  In the self-consistent solution,
  $r \simeq 0.259$ and $\ell \simeq 6.334$.}
   \label{Fig:strong_spec}
\end{figure}

The corresponding quasiparticle band structure is shown in Fig.~\ref{Fig:strong_spec}. Nodal points exist at the Fermi energy, in the bands for which the pseudospin (defined earlier) has an eigenvalue $-D$. They occur at the wave vectors ${\bf k}_W$, determined in terms of the hybridized bands,
\begin{eqnarray}
\nonumber \mathcal{E}^{(+,+)}_{-D}({\bf k}_W)= \mathcal{E}^{(-,+)}_{-D}({\bf k}_W),\\
\mathcal{E}^{(+,-)}_{-D} ({\bf k}_W)= \mathcal{E}^{(-,-)}_{-D}({\bf k}_W),
\end{eqnarray}
for the upper and lower branches, respectively. The Weyl nodes appear along the Z lines (lines connecting the X and W points) in the three planes of the BZ as illustrated in Fig.~\ref{Fig:3D_BZ}. 
This is specific to the zinc blende lattice. For other types of lattices, the Weyl nodes may occur away from the high symmetry parts of the 
BZ. 

We note that the bands near the Fermi energy 
have a width much reduced from the noninteracting value. This width is given by the Kondo energy.
Compared to the bare width of the conduction electron band, the reduction factor corresponds to $r^2$
[which is about 0.067 in the specific case shown in Fig.~\ref{Fig:strong_spec}].
We remark that, in the absence of the hybridization between the $f$- and conduction electrons, the ground state would be an insulator 
instead of a semimetal: the $f$-electrons would be half-filled and form a Mott insulator, while the conduction electrons would be empty, 
forming a band insulator. All these imply that the nodal excitations develop out of the Kondo effect.
%%%%%%%%%%%%%
%%%%%%%%%%%%%

\begin{figure}[t]
   \centering
     \subfigure[]{\label{Fig:3d_sc}\includegraphics[width= 1.5 in]{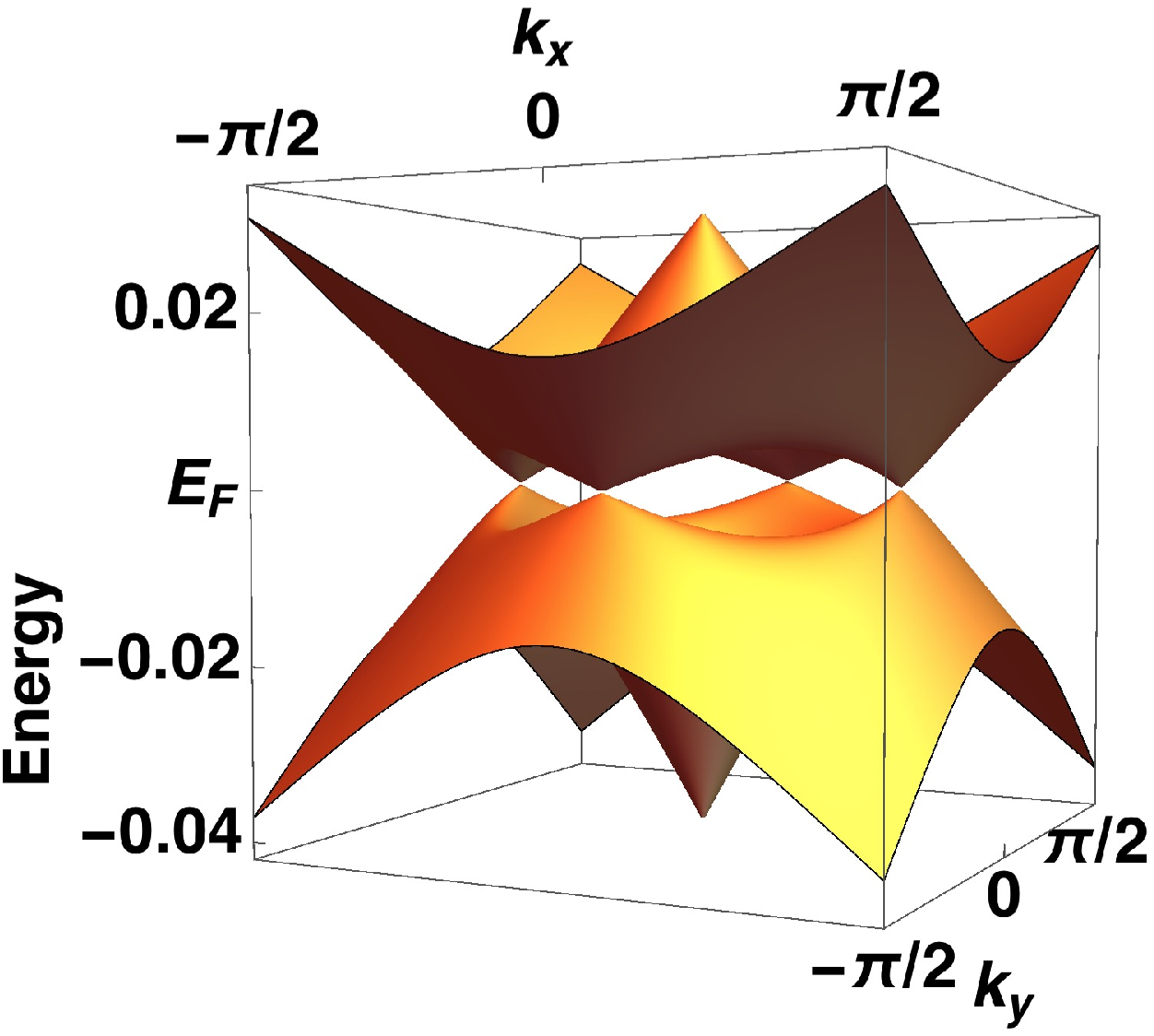}}
     \subfigure[]{\label{Fig:bulk_berry}\includegraphics[width= 1.5 in]{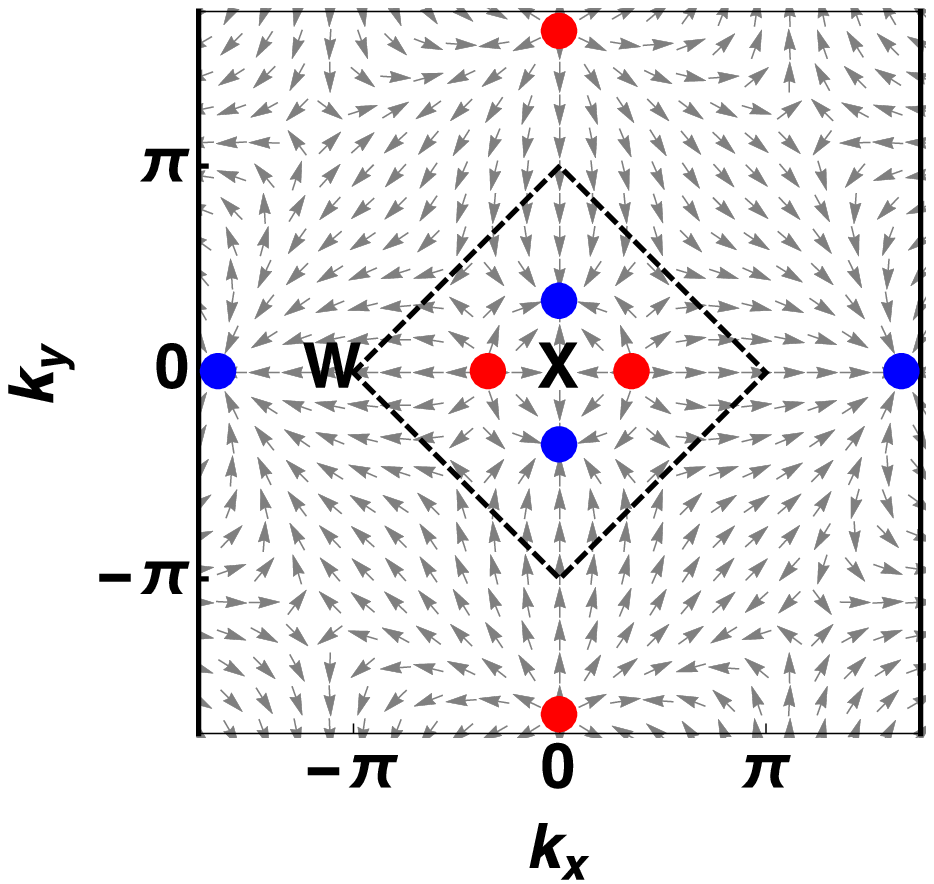}}
    \caption{{\bf Characterization of the Weyl nodes.} 
   The plots are in the $(k_x, k_y)$-plane of the four Weyl nodes at $k_z=2\pi$ (gray plane in Fig.~\ref{Fig:3D_BZ}).
    {\bf (a),} 
    Energy dispersion, showing the band degeneracies at the Weyl node points and  a strong reduction of the bandwidth;
    {\bf (b),} 
    The distribution of the Berry curvature field.
    The bare parameters are the same as 
in Fig.~\ref{Fig:strong_spec}. 
    }
\label{Fig:3d_dispersion}
\end{figure}

To demonstrate the monopole flux structure of the Weyl nodes, we calculate the Berry curvature in the strong coupling regime. We show the results at the $k_z = 2\pi$ boundary of the 3D BZ, in the grey plane of Fig.~\ref{Fig:3D_BZ} whose dispersion is shown in Fig.~\ref{Fig:bulk_berry}. In Fig.~\ref{Fig:bulk_berry}, the arrows represent the field's unit-length 2D projection onto the $k_xk_y$-plane, $\hat{\Omega}(k_x,k_y,2\pi) = |\vec{\Omega}(k_x,k_y,2\pi)|^{-1} \begin{pmatrix}\Omega_{yz}(k_x,k_y,2\pi),~\Omega_{zx}(k_x,k_y,2\pi)\end{pmatrix}$. The Weyl node locations (blue/red circles) are clearly indicated by the arrows flowing in or out, representing negative or positive monopole ``charge.'' 

We next analyze the surface states. Focusing on the $(001)$ surface,
we find the following energy dispersion for the surface states:
\begin{eqnarray}
\nonumber &&\hspace{-0.7cm} \mathfrak{E}(k_x, k_y) 
=  - 2\sin\left(\frac{k_x}{4}\right) \sin\left(\frac{k_y}{4}\right) + \frac{V_s^2+(E_s)^2}{2E_s} \hspace{2cm} \\
&& -  \sqrt{ \left( 2\sin\left(\frac{k_x}{4}\right)\sin\left(\frac{k_y}{4}\right)-\frac{V_s^2-(E_s)^2}{2E_s} \right)^2 + V_s^2},
\label{eq:ss_spectrum}
\end{eqnarray}
where we define the parameters $(V_s, E_s, \mu_s) = (r V, E_d + \ell, -(rV)^2/(E_d + \ell))$. (For the derivation, see Supplementary Information.) In Figs.~\ref{Fig:surf_1d_path}-\ref{Fig:surf_1d_sc}, we show the energy dispersion along a high symmetry path in the ${\bf k}$-space. The solid lines represent the surface states, and the dashed lines show where they merge with the bulk states and can no longer be sharply distinguished. The surface electron spectrum has a width that is similarly narrow as the bulk electron band (compare Fig.~\ref{Fig:surf_1d_sc} with Fig.~\ref{Fig:3d_sc}), implying that the surface states also come from the Kondo effect. The surface Fermi arcs (where $\mathfrak{E}(k_x, k_y) = 0$) connect the Weyl nodes along $k_x = 0$ and $k_y = 0$, separating the positive and negative energy surface patches, marked by the solid black lines in Fig.~\ref{Fig:surf_1d_path}. 

As is typical for strongly correlated systems, the most dominant interactions in heavy fermion systems are onsite, making it important to study them in lattice models (as opposed to the continuum limit). Our explicit calculations have been possible using a well-defined model on a diamond lattice that permits inversion-symmetry breaking. 
Nonetheless, we expect our conclusion to qualitatively apply to other noncentrosymmetric 3D systems. Finally, the WKSM is expected to survive the effect of a time-reversal symmetry breaking term, such as a magnetic field; this is illustrated in the Supplementary Information.
%%%%%%%%%%%
%%%%%%%%%%%

We now turn to the implications of our results for heavy fermion semimetals. The entropy from the bulk Weyl nodes will be dictated by the velocity $v^*$, and the corresponding specific heat per unit volume has the following form (see Supplementary Information):
\begin{eqnarray}
c_v \sim (k_B T/\hbar v^*)^3 k_B .
\label{Eq:entropy}
\end{eqnarray}
We stress that this expression is robust against the residual interactions of the nodal excitations (see Supplementary Information).
The utility of thermodynamical quantities as a key signature reflects an important distinction of the WKSM from weakly correlated Weyl semimetals. The Kondo temperature of typical heavy fermion systems is 
considerably smaller than the Debye temperature. This is to be contrasted with the weakly correlated systems, in which the bandwidth of the conduction electrons is 
much larger than the Debye temperature. Therefore, in a WKSM, the nodal contributions to the entropy would dominate over the phonon component. The corresponding form of entropy also implies that the nodal excitations will have large contributions to the thermopower.

Eq.~(\ref{Eq:entropy}) can be readily tested, given that 
there is 
a considerable number of semimetallic heavy fermion compounds~\cite{SiPaschen2013}. A noncentrosymmetric heavy-fermion system Ce$_3$Bi$_4$Pd$_3$ has recently been discovered to display semimetal behavior based on transport measurements, and its specific heat is well described in terms  of Eq.~(\ref{Eq:entropy})~[Ref.~\cite{Dzsaber2016}]. A fit in terms of our theoretical expression Eq.~(S29) (Supplementary Information) reveals an effective velocity, $v^*$, that is three orders of magnitude smaller than that expected for weakly correlated systems, reflecting the reduction in the energy scale -- the Kondo temperature for Ce$_3$Bi$_4$Pd$_3$ -- from the bandwidth of the latter by a similar order of magnitude~\cite{Dzsaber2016}. This analysis provides strong evidence 
that Ce$_3$Bi$_4$Pd$_3$ is a candidate WKSM system with strongly-correlated Weyl nodes, and provides the motivation for further studies on such quantities as magnetotransport and high-resolution angle-resolved photoemission spectroscopy (ARPES) in this system. More recently, Eq.~(\ref{Eq:entropy}) has been used to fit the specific heat of another heavy fermion system, YbPtBi, suggesting it be another candidate WKSM system \cite{Guo2017}.
%%%%%%%%%%%%%%%%%%%%%%%%%%%%%%%%%

Our theoretical results provide guidance in the search for Weyl semimetals in other heavy fermion systems. For instance, in the $4f$-based system CeSb, Weyl physics has been suggested based on magnetotransport measurements ~\cite{Guo1611,Alidoust2016}. Even though any nodes in this system are likely away from the Fermi energy, the fact that it is a Kondo system with low energy scales leads to the expectation
that they can be tuned towards the Fermi energy by pressure or chemical doping, and we propose specific heat measurements and our ~\eqref{Eq:entropy} as a means of ascertaining the role of the $4f$-electrons in this system. In addition, the noncentrosymmetric CeRu$_4$Sn$_6$ also displays semimetal properties~\cite{Guritanu2013} and has been discussed as a potential topological system~\cite{Sunderman2015}. Its electronic structure has been studied by {\it ab initio} calculations combined with dynamical mean field theory (DMFT)~\cite{Guritanu2013,Wissgott2016} or the Gutzwiller projection method~\cite{Yu2016}. While  the two types of calculations disagree on the low-energy dispersion and the latter study does not appear to capture the strong renormalizations expected in a Kondo system, the existence of linearly-dispersing nodes and their Weyl nature have been suggested in the latter study. The low-temperature specific heat in single crystalline CeRu$_4$Sn$_6$~\cite{Paschen2010} implies the importance of the $4f$-electrons to the low-energy physics but does not appear to have the form of ~\eqref{Eq:entropy}. Our theoretical results suggest that further thermodynamic and thermoelectrical studies will be instructive in ascertaining the potential WKSM nature of CeRu$_4$Sn$_6$.
%%%%%%%%%%%%%%
\begin{figure}[t]
   \centering
      \subfigure[]{\label{Fig:surf_1d_path}\includegraphics[width= 1.5 in]{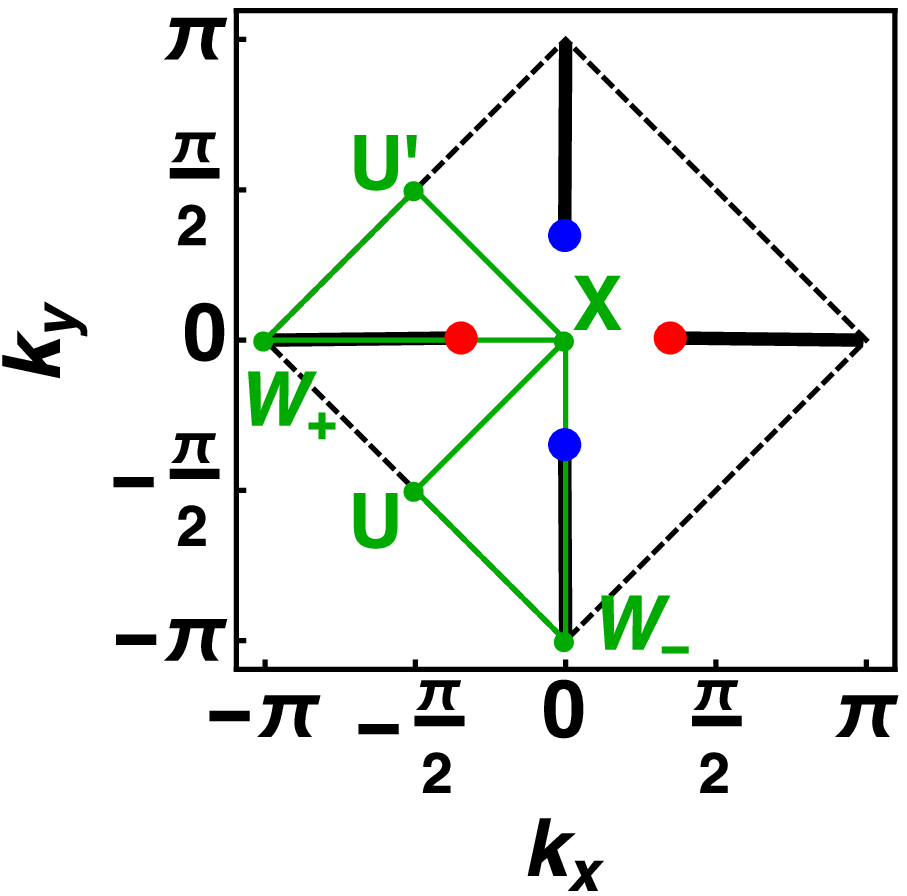}}~~
     \subfigure[]{\label{Fig:surf_1d_sc}\includegraphics[width= 1.5 in]{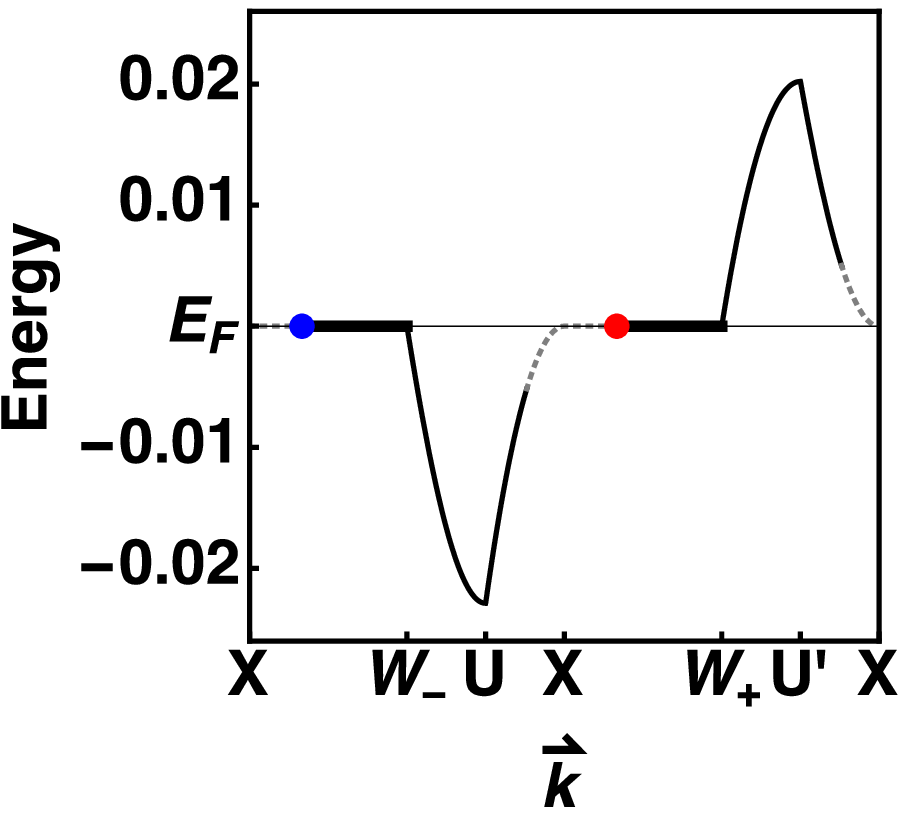}}
    \caption{{\bf Energy dispersion of the surface electronic states.}
   The spectrum shows the (anti-)Weyl nodes marked with blue (red). 
   The solid black lines connecting the nodes represent Fermi arcs and 
   black dashed lines represent the BZ around the $X$ point. 
   {\bf (a),} 
    High-symmetry ${\bf k}$-space contour taken on the BZ boundary at $k_z=2\pi$;
    {\bf (b),}  
    Energy dispersion of the surface state 
    along the path specified in (a); the grey dotted line denotes the decay of 
    the surface states into the bulk states. 
    The parameters are the same as in Fig.~\ref{Fig:strong_spec}.
}
\label{Fig:berry_surface}
\end{figure}
%%%%%%%%%%%%%%

We close with several observations. First, we have focused on a model defined on an nonsymmorphic diamond lattice,
in which, for non-interacting electrons,  the crystallographic space group symmetry allows for a filling-enforced 
semimetal state~\cite{Kane_3ddirac,AMV_RMP}. 
Our study here demonstrates that for Kondo systems defined on such a lattice and in the presence 
of inversion-symmetry breaking, the 
WKSM phase arises in a robust way. 
On the other hand, for non-interacting systems in 3D a Dirac semimetal can also arise at the phase transition between 
topologically distinct insulating states.
It will therefore be instructive to search for the WKSM
state at topological phase transitions in Kondo systems; we leave this matter for 
future studies.

Second, our work provides a proof of principle demonstration for the emergence of a WKSM
phase in a Kondo lattice with inversion symmetry breaking. This makes it likely that
such a phase arises in inversion-symmetry-breaking 3D Kondo lattices with other crystallographic symmetries.
The candidate WKSM material Ce$_3$Bi$_4$Pd$_3$ has a nonsymmorphic space group (220), and its Kondo-driven Weyl nodes may very well be enforced by its crystallographic symmetry and electron filling. Our findings here motivates further studies that incorporates the realistic electronic structure of Ce$_3$Bi$_4$Pd$_3$.

Third, in the WKSM state advanced here, the electron correlations produce a zeroth-order effect given that the localized moments of the $4f$-electrons underlie the Weyl excitations. Correspondingly, the renormalization factors (for the nodal velocity)
are extremely
large, typically on the order of $10^2-10^3$. This distinguishes the WKSM from other types of interacting Weyl semimetals discussed previously. The large renormalization factor is responsible for the possibility of using thermodynamics to probe the Weyl nodes. Our work also sets the stage for calculations of additional signature properties for the Weyl physics, such as the optical conductivity and related dynamical quantities.

Fourth, the context of heavy fermion systems links the WKSM phase advanced here with the topological Kondo insulators~\cite{Dzero2010} and the physics of their surface states~\cite{Nikolic2014}. Nonetheless, the WKSM as a state of matter is distinct: in the bulk, it features strongly renormalized Weyl nodes instead of being fully gapped; on the surface, it hosts Fermi arcs from a band with a width of the Kondo energy, which are to be contrasted with the surface Dirac nodes of the topological Kondo insulators. This also dictates that the WKSM will be realized in heavy fermion materials that are quite different from those for the topological Kondo insulators, as exemplified by the aforementioned Ce$_3$Bi$_4$Pd$_3$.

Finally, the emergence of a WKSM makes it natural for quantum phase
transitions in heavy fermion systems from such a topological semimetal
to magnetically ordered and other correlated paramagnetic
states. Moreover, the existence of Weyl nodes also enhances the effect of long-range Coulomb interactions. While the density fluctuations of the $f$-electrons are strongly suppressed in the local-moment regime, it still will be instructive to explore additional nearby phases such as charge-density-wave order \cite{Wei2014};
other type of long-range interactions could produce topologically-nontrivial Mott insulators \cite{MorimotoNagaosa2016}. As such, the study of WKSM and related semimetals promises to shed new light on the global phase diagram of quantum critical heavy fermion systems~\cite{SiPaschen2013} and other strongly correlated materials~\cite{Krempa2014}.

In summary, we have demonstrated an emergent Weyl-Kondo semimetal phase in a model of heavy fermion systems 
with broken inversion symmetry, and have determined the surface electronic spectra which reveal Fermi arcs. 
The nodal excitations of the WKSM phase develop out of the Kondo effect. 
This leads to unique experimental signatures for such a phase, which are realized in 
a 
noncentrosymmetric heavy fermion system.
Our results are expected to guide the experimental search for $f$-electron-based Weyl semimetals. In general, they open the door for studying topological semimetals in the overall context of quantum phases and their transitions in strongly correlated electron systems and, conversely, broaden the reach of strongly correlated gapless and quantum critical states of matter.

%%%%%%%%%%%%%%
\matmethods{
\subsection*{The 3D periodic Anderson model} 
The three terms in the model of Eq.~(\ref{Eq:H_pam}) are presented here in more detail. The strongly correlated $d$-electrons are specified by
\begin{eqnarray}
H_d=E_{d} \sum_{i, \sigma} d_{i \sigma}^\dagger d_{i \sigma} + U \sum_i n^d_{i \up} n^d_{i \dn}.
\label{eq:hd}
\end{eqnarray}
The hybridization term is as follows:
\begin{eqnarray}
H_{cd} =
V \sum_{i,\sigma} \left( d_{i \sigma}^\dagger c_{i \sigma} + \Hc\right) .
\label{eq:hybridization}
\end{eqnarray}
In the above two equations, the site labeling $i$ means $i = (\bm{r}, a)$, where $\bm{r}$ runs over the Bravais lattice of unit cells and $a$ runs over the two sites, $a=A,~B$, in the unit cell.

The conduction electron Hamiltonian $H_c$ realizes a modified Fu-Kane-Mele model~\cite{FKMmodel07}, and 
is expressed as $H_c=\sum_{\bf k}\Psi^\dagger_{\bf k}~h_{\bf k}~\Psi_{\bf k}$,
where 
$\Psi^T_{\bf k}=\begin{pmatrix} c_{{\bf k}\up,A}& c_{{\bf k}\up,B} & c_{{\bf k}\dn,A} & c_{{\bf k}\dn,B}\end{pmatrix}$,
and 
\begin{eqnarray}
h_{\bf k}=&& \sigma_0\left( u_1({\bf k})\tau_x + u_2({\bf k})\tau_y + m\tau_z\right) + \lambda \left( {\bf D({\bf k})} \cdot \bm{ \sigma}\right) \tau_z .\label{eq:hc_sigma}
\end{eqnarray}
Here, $\bm{\sigma} = (\sigma_x,\sigma_y,\sigma_z)$ and $\bm{\tau} =(\tau_x,\tau_y,\tau_z)$ are the Pauli matrices acting on the spin and sublattice spaces, respectively, and $\sigma_0$ is the identity matrix. In the first term, $u_1({\bf k})$ and $u_2({\bf k})$ are determined by the conduction electron hopping, $t_{\la ij \ra } = t$ between nearest-neighbor sites ($\la ij \ra $). The second term specifies a Dresselhaus-type spin-orbit coupling between the second-nearest-neighbor sites ($\la \la ij \ra \ra$), which is of strength $\lambda$ and involves vector ${\bf D}({\bf k})=\begin{pmatrix}D_x({\bf k}),D_y({\bf k}),D_z({\bf k})\end{pmatrix}$.
Specifically,
\begin{eqnarray}
&& u_1({\bf k})= t\left(1+\sum_{n=1}^3\cos({\bf k}\cdot {\bf a}_n)\right),\label{eq:d1}\\
&& u_2({\bf k})= t\sum_{n=1}^3 \sin({\bf k}\cdot{\bf a}_n ),\label{eq:d2}\\
\nonumber && {D_x({\bf k})} = \sin({\bf k}\cdot {\bf a}_2 ) - \sin({\bf k}\cdot {\bf a}_3) - \sin({\bf k}\cdot ({\bf a}_2 - {\bf a}_1 )) \\
&& \hspace{1.2cm} + \sin({\bf k}\cdot ({\bf a}_3 -{\bf a}_1)), \label{eq:Dx}
\end{eqnarray}
and $D_y,~D_z$ are obtained by permuting the fcc primitive lattice vectors ${\bf a}_n$.
The canonical (unitary) transformation, $\breve{\Psi}_{\bf k} = S^\dagger_\sigma\Psi_{\bf k}$, leads to
\begin{eqnarray}
&& H_c = \sum_{\bf k} \breve{\Psi}^\dagger_{\bf k} \begin{pmatrix} 
 h_{{\bf k} +} & 0 \\ 0 & h_{{\bf k}-} 
\end{pmatrix} 
	\breve{\Psi}_{\bf k},\\
&& h_{{\bf k} \pm} = u_1({\bf k})\tau_x + u_2({\bf k})\tau_y + ( m \pm \lambda D({\bf k}))\tau_z.
\end{eqnarray}
We have used a pseudospin basis~\cite{Ojanen13}, defined by the eigenstates  $|\pm D\ra$ with eigenvalues 
\begin{eqnarray}
\frac{{\bf D}\cdot \bm{\sigma}}{D}|\pm D\ra=\pm|\pm D\ra
 \label{Eq:pseudospin}
\end{eqnarray}
where $D({\bf k}) \equiv |{\bf D({\bf k})}|$.
The eigenenergies of the $|\pm D\ra$ sectors are simply obtained to be
\begin{eqnarray}
\varepsilon^{\tau}_{\pm D} =\tau\sqrt{u_1({\bf k})^2+u_2({\bf k})^2+(m\pm\lambda D({\bf k}))^2}
\end{eqnarray}
where $\tau=(+,-)$. We use this transformation on the full Anderson model in the strong coupling limit, at the saddle point level where the Lagrange multiplier $\ell_i$, which enforces the local constraint $b_i^\dagger b_i + \sum_\sigma f^\dagger_{i\sigma} f_{i\sigma} = 1$, takes a uniform value,
  $\ell$. This corresponds to $\breve{\Xi}_{\bf k} = S_\sigma^\dagger \Xi_{\bf k}$. Anticipating the separability of the $|\pm D\ra$ sectors 
 by specifying $\breve{\Psi}^T_{\bf k} = \begin{pmatrix} \breve{\psi}^T_{{\bf k}+}, \breve{\psi}^T_{{\bf k} - } \end{pmatrix}$,
 $\breve{\Xi}^T_{\bf k} = \begin{pmatrix} \breve{\xi}^T_{{\bf k}+}, \breve{\xi}^T_{{\bf k} - } \end{pmatrix}$, where
$\breve{\psi}_{\bf k\pm}^T = \begin{pmatrix} \breve{\psi}_{{\bf k}\pm ,A}, \breve{\psi}_{{\bf k}\pm,B} \end{pmatrix}$ and
$\breve{\xi}_{\bf k\pm}^T = \begin{pmatrix} \breve{\xi}_{{\bf k}\pm ,A},\breve{\xi}_{{\bf k}\pm ,B} \end{pmatrix}$, we obtain the strong coupling Hamiltonian
\begin{eqnarray}\label{Eq:H_sc}
H^{s} = \sum_{{\bf k}, a=\pm} 
	\begin{pmatrix} 
		\breve{\psi}^\dagger_{{\bf k}a} & \breve{\xi}^\dagger_{{\bf k}a} 
	\end{pmatrix} 
\begin{pmatrix} 
	h_{{\bf k}a} - \mu \mathbbm{1}_2 & r V \mathbbm{1}_2 \\
	rV \mathbbm{1}_2 & (E_d + \ell) \mathbbm{1}_2 
\end{pmatrix} 
\begin{pmatrix} 
	\breve{\psi}_{{\bf k}a} \\ \breve{\xi}_{{\bf k}a}
\end{pmatrix},
\end{eqnarray}
which separates as $H^{s} = \mathcal{H}^{s}_+ + \mathcal{H}^{s}_-$. We obtain the full spectra of the eight hybridized bands,
\begin{eqnarray}
\mathcal{E}^{(\tau,\alpha)}_{\pm D}({\bf k})= \frac{1}{2} \left[ E_s + \tilde{\varepsilon}^{\tau}_{\pm D}  +\alpha \sqrt{\left( E_s - \tilde{\varepsilon}^{\tau}_{\pm D}  \right)^2 + 4V_s^2}\right],\label{Eq:hb_spec}
\end{eqnarray}
where $\alpha=(+,-)$ indexes the upper/lower quartet of bands, $\tilde{\varepsilon}^\tau_{\pm D}=\varepsilon^\tau_{\pm D}-\mu$, 
and $(E_s,~V_s)= (E_d+\ell, r V)$. 
In the Supplemental Information,
we prove that the $|+D\ra$ sector is always gapped, whereas the $|-D\ra$ sector allows Weyl nodes when $0<\frac{m}{4|\lambda|} < 1$, and determine $\mu=-V_s^2/E_s$ which fixes the Fermi energy at the Weyl nodes.

To determine $r,~\ell$, $H^{s}$ must be solved self-consistently from the 
saddle-point equations
\begin{eqnarray}
\nonumber &&	\frac{1}{2 N_u} \sum_{{\bf k},a=\pm} \left\la \breve{\xi}^\dagger_{{\bf k}a} \breve{\xi}_{{\bf k}a}\right\ra +r^2 = 1, \\
&&	\frac{V}{4N_u} \sum_{{\bf k},a=\pm} \left[  \left\la \breve{\psi}^\dagger_{{\bf k}a} \breve{\xi}_{{\bf k}a}\right\ra + \Hc\right] + r \ell = 0,
\end{eqnarray}
where $N_u$ is the number of the unit cell. The equations are solved  on a $64 \times 64 \times 64$ cell of the diamond lattice, with error $\epsilon\leq\mathcal{O}(10^{-5})$.
}

%%%%%%%%%
\showmatmethods % Display the Materials and Methods section

\acknow{We acknowledge useful discussions with J. Analytis, S. Dzsaber, M. Foster, D. Natelson, A. Prokofiev, B. Roy, F. Steglich,
S. Wirth,  and H. Q. Yuan. This work has been supported in part by the NSF Grant No.\ DMR-1611392 (H.-H.L. and Q.S.),
the ARO Grant No.\ W911NF-14-1-0525 (S.G.), the Robert A.\ Welch Foundation Grant No.\ C-1411 (S.G. and Q.S.), the NSF Grant No.\ DMR-1350237 and a Smalley Postdoctoral Fellowship of Rice Center for Quantum Materials (H-H. L.), the ARO grant W911NF-14-1-0496 and  the Austrian Science Funds FWF I2535-N27 (S.P.). 
}

\showacknow % Display the acknowledgments section

% \pnasbreak splits and balances the columns before the references.
% If you see unexpected formatting errors, try commenting out this line
% as it can run into problems with floats and footnotes on the final page.
%\pnasbreak

% Bibliography
\bibliography{biblio5KWSM}

\end{document}

\verticaladjustment{-2pt}

\maketitle
\thispagestyle{firststyle}
\ifthenelse{\boolean{shortarticle}}{\ifthenelse{\boolean{singlecolumn}}}{}

%%%%%%%%%%%%%%%%%%%
\section*{Analysis of the bulk spectrum}
In the strong coupling regime that is relevant to heavy fermion systems, 
we can approach the prohibition of $d_\sigma$ fermion double occupancy by an auxiliary boson method~\cite{Hewson_Kondo_book}.
Representing $d^\dagger_{i\sigma} = f^\dagger_{i\sigma} b_{i}$, the $f^\dagger_{i\sigma}$ ($b_i$) are fermonic (bosonic) operators satisfying the constraint $b_i^\dagger b_i + \sum_\sigma f^\dagger_{i\sigma} f_{i\sigma} = 1$.
At the saddle point level, we replace $b^\dagger_i,~b_i \rightarrow r$, and introduce a Lagrange multiplier $\ell$ to enforce the local constraint. Defining 
$\Xi_{\bf k}^T \equiv \begin{pmatrix} d_{{\bf k}\up,A}& d_{{\bf k}\up,B}& d_{{\bf k}\dn,A}& d_{{\bf k}\dn,B}\end{pmatrix}$ and $\Psi^T_{\bf k} \equiv \begin{pmatrix} c_{{\bf k}\uparrow, A} & c_{{\bf k}\uparrow, B} & c_{{\bf k}\downarrow,A} & d_{{\bf k}\downarrow, B} \end{pmatrix}$, we can transform Hamiltonian $H_c$, $H_{cd}$, and $H_d$ into the pseudospin basis using $\breve{\Xi}_{\bf k} = S_\sigma^\dagger \Xi_{\bf k}$ and $\breve{\Psi}_{\bf k} = S_\sigma^\dagger \Psi_{\bf k}$, with $S^\dagger_\sigma=U_\sigma\tau_0$ being a unitary matrix which consists of the $|\pm D\ra$ eigenvectors. The effective hybridization becomes $ r V$, which  is nonzero only for $V>V_c$, 
whenever the conduction-electron density of states 
has a pseudogap form near the Fermi energy~\cite{Feng13}.
The hybridization part can be re-expressed as
\begin{eqnarray}
H_{cd} = \sum_{\bf k} \left[ \breve{\Psi}_{\bf k}^\dagger \cdot rV \mathbbm{1}_4 \cdot \breve{\Xi}_{\bf k} + \Hc\right].
\end{eqnarray}
Introducing
$\breve{\Psi}^T_{\bf k} = \begin{pmatrix} \breve{\psi}^T_{{\bf k}+}& \breve{\psi}^T_{{\bf k} - } \end{pmatrix}$,
 $\breve{\Xi}^T_{\bf k} = \begin{pmatrix} \breve{\xi}^T_{{\bf k}+}& \breve{\xi}^T_{{\bf k} - } \end{pmatrix}$, where
$\breve{\psi}_{\bf k\pm}^T = \begin{pmatrix} \breve{\psi}_{{\bf k}\pm ,A}& \breve{\psi}_{{\bf k}\pm,B} \end{pmatrix}$ and
$\breve{\xi}_{\bf k\pm}^T = \begin{pmatrix} \breve{\xi}_{{\bf k}\pm ,A}& \breve{\xi}_{{\bf k}\pm ,B} \end{pmatrix}$, we find that the strong-coupling Hamiltonian can straightforwardly be written in the $|\pm D\ra$ basis as $H^{s} = \sum_{a = \pm} \mathcal{H}^{s}_a$,
\begin{eqnarray}
\mathcal{H}^{s}_a= &&\sum_{{\bf k}, a=\pm}  \begin{pmatrix} \breve{\psi}_{{\bf k}a}^\dagger & \breve{\xi}_{{\bf k}a}^\dagger \end{pmatrix} 
	\begin{pmatrix} 
		h_{{\bf k}a} - \mu \mathbbm{1}_2 & V_s \mathbbm{1}_2 \\
		V_s \mathbbm{1}_2 & E_s \mathbbm{1}_2 
	\end{pmatrix} 
	\begin{pmatrix} 
		\breve{\psi}_{{\bf k}a} \\ \breve{\xi}_{{\bf k}a}
	\end{pmatrix},\label{Seq:scpm}
\end{eqnarray}
where $V_s \equiv rV$ and $ E_s \equiv E_d + \ell$. Straightforward diagonalization of the strong-coupling Hamiltonian yields a set of four quasiparticle bands for each spin sector as
\begin{eqnarray}
\mathcal{E}^{(\tau,\alpha)}_{\pm D}({\bf k})&&= \frac{1}{2} \left[ E_s + \tilde{\varepsilon}^{\tau}_{\pm D}  +\alpha \sqrt{\left( E_s - \tilde{\varepsilon}^{\tau}_{\pm D}  \right)^2 + 4 V_s^2}\right],\\
\tilde{\varepsilon}^\tau_{\pm D}&&=\varepsilon^\tau_{\pm D}-\mu,\\
\varepsilon^{\tau}_{\pm D} &&=\tau\sqrt{u_1({\bf k})^2+u_2({\bf k})^2+(m\pm\lambda D({\bf k}))^2},
\end{eqnarray}
where $\tau=(+,-)$, and $\alpha=(+,-)$ indexes the upper/lower quartet of bands, respectively.

To gain a deeper understanding of the gap structure of the hybridized bands, it is more convenient to first diagonalize the conduction electron part of the Hamiltonian, which is possible since the off-diagonal blocks and the bottom right block are all proportional to $2\times2$ identity matrices. 
Diagonalizing the conduction electron part of the Hamiltonian, we can rewrite the strong coupling Hamiltonian in a diagonal form,
\begin{eqnarray}
h^D_{{\bf k}\pm} = 
\begin{pmatrix} \varepsilon^+_{\pm D}& 0 \\
0 & \varepsilon^-_{\pm D}
\end{pmatrix},
\end{eqnarray}
and in the new basis, the Hamiltonian becomes
\begin{eqnarray}
\mathcal{H}^{s}_\pm  &=& \sum_{\bf k} \begin{pmatrix} (\psi^D_{{\bf k}\pm})^\dagger & (\xi^D_{{\bf k}\pm})^\dagger \end{pmatrix} 
\begin{pmatrix} 
h_{{\bf k}a}^D - \mu \mathbbm{1}_2 & V_s \mathbbm{1}_2 \\
V_s \mathbbm{1}_2 & E_s \mathbbm{1}_2 
\end{pmatrix} 
\begin{pmatrix} 
\psi^D_{{\bf k}\pm} \\ \xi^D_{{\bf k}\pm}
\end{pmatrix}.\label{Seq:wcpm}
\end{eqnarray}
We can then directly see that the matrix elements associated with the 1st and 3rd fields are decoupled from the 2nd and 4th fields, which means we can simplify the $4\times4$ matrix in either $|\pm D\ra$ sector, to be two $2\times2$ matrices, which allows us to examine the eigenenergy bands analytically. 
Below, we discuss the cases in different $| \pm D \ra$ sectors separately. 
Our main conclusion below is that the Weyl-Kondo semimetal phase can only occur at the $| - D \ra$ sector in the hybridized band regime, 
and the hybridized bands in $| + D\ra$ sector \textit{always} remain gapped.
\\

%%%%%%%%%%%%%%%%%
\centerline{ $|+D\ra$ sector:}
For further analysis in the band gaps, we assume that $E_s$ lies well below the conduction electron bands $ \varepsilon^{\tau}_{\pm D}$. In addition, for the condition for $1/4$-filling, $E_s$ is required to be positive $E_s >0$. Focusing on the two $2\times2$ matrices of the $|+D\ra$ sector, we can separate the Hamiltonian into $\mathcal{H}^{s}_+ = \sum_{\alpha,\tau} H^{s,\tau}_{+,\alpha}$, the energies obtained are
\begin{eqnarray}
&& \mathcal{E}^{(+,\alpha)}_{+D} = \frac{1}{2} \left[ E_s + \tilde{\varepsilon}^+_{+D}  +\alpha \sqrt{\left( E_s - \tilde{\varepsilon}^+_{+D}  \right)^2 + 4 V_s^2}\right],\\
&& \mathcal{E}^{(-,\alpha)}_{+D} = \frac{1}{2} \left[ E_s + \tilde{\varepsilon}^-_{+D}  +\alpha \sqrt{\left( E_s -  \tilde{\varepsilon}^-_{+D} \right)^2 + 4 V_s^2}\right].
\end{eqnarray}
Since $\varepsilon^+_{+D}  > \varepsilon^-_{+D}$, we can see $\mathcal{E}^{(+,+)}_{+D} > \mathcal{E}^{(-,-)}_{+D}$ and these two bands always remain gapped. Similarly, within each pair of branches $\mathcal{E}^{(\tau,+)}_{+D}> \mathcal{E}^{(\tau,-)}_{+D}$, and they should be always gapped. The only possibility that the gap closes occurs between $\mathcal{E}^{(+,-)}_{+D}$ and $\mathcal{E}^{(-,+)}_{+D}$. If there is a crossing between them at some momenta ${\bf k} = {\bf k}_0$, the two bands should be degenerate  $\mathcal{E}^{(+,-)}_{+D}({\bf k}_0) = \mathcal{E}^{(-,+)}_{+D}({\bf k}_0)$, which leads to
\begin{eqnarray}
\nonumber \varepsilon^+_{+D} ({\bf k}_0) - \varepsilon^-_{+D}({\bf k}_0)  &=& \sqrt{\left( E_s - \tilde{\varepsilon}^+_{+D} ({\bf k}_0)  \right)^2 + 4 V_s^2} + \sqrt{\left( E_s -  \tilde{\varepsilon}^-_{+D} ({\bf k}_0)\right)^2 + 4 V_s^2} \\
\nonumber &\geq& \varepsilon^+_{+D} ({\bf k}_0) + \varepsilon^-_{+D} ({\bf k}_0)- 2 E_s \\
&\Rightarrow & \varepsilon^-_{+D} ({\bf k}_0) \leq E_s,
\end{eqnarray}
where we use the assumption that $E_s < \epsilon^\tau_{\pm D}$ in the second line. This in turn leads to a contradiction to our initial condition that the $d$ fermion Fermi energy is well below the four conduction electron bands, $E_s < \varepsilon^\tau_{\pm D} $. Therefore, we conclude that there cannot be any crossing between $\mathcal{E}^{(+,-)}_{+D}$  and $\mathcal{E}^{(-,+)}_{+ D}$. The hybridized bands in the $| + D \ra$ sector \textit{always} remain gapped at any momenta and Weyl nodes cannot reside there. Now let's examine the $|-D\ra$ sector.
\\

%%%%%%%%%%%%%%%%%%%%%%%%%
\centerline{$|-D\ra$ sector:}

In the $|-D\ra$ sector, we can also decompose the $4\times4$ Hamiltonian matrix to two $2\times 2$ matrices, $\mathcal{H}^{s}_+ = \sum_{\alpha,\tau} H^{s,\tau}_{+,\alpha}$, which gives the eigenvalues as
\begin{eqnarray}
&& \mathcal{E}^{(+,\alpha)}_{-D} = \frac{1}{2} \left[ E_s + \tilde{\varepsilon}^+_{-D}  +\alpha \sqrt{\left( E_s - \tilde{\varepsilon}^+_{-D}  \right)^2 + 4 V_s^2}\right],\\
&& \mathcal{E}^{(-,\alpha)}_{-D} = \frac{1}{2} \left[ E_s + \tilde{\varepsilon}^-_{-D} +\alpha  \sqrt{\left( E_s -  \tilde{\varepsilon}^-_{-D} \right)^2 + 4 V_s^2}\right].
\end{eqnarray}
The bands' dispersions associated with the conduction electrons show Weyl nodes at certain momenta, i.e. ${\bf k} = {\bf k}_W$, where $\varepsilon^+_{-D} ({\bf k}_W) = \varepsilon^{-}_{-D}({\bf k}_W) = 0$. 
There are actually 12 inequivalent ${\bf k}_W$ along the $X-W$ lines on the 3D BZ boundary, determined by the condition $\frac{m}{4|\lambda|}=\sin(\frac{k_0}{2})$, where ${\bf k}_W$ is a cyclic permutation of the elements in a vector $(k_0,0,\pm2\pi)$.
At ${\bf k} = {\bf k}_W$, in the hybridized bands we then have
\begin{eqnarray}
&& \mathcal{E}^{(+,+)}_{-D}({\bf k}_W)= \mathcal{E}^{(-,+)}_{-D}({\bf k}_W),\\
&& \mathcal{E}^{(+,-)}_{-D}({\bf k}_W) = \mathcal{E}^{(-,-)}_{-D}({\bf k}_W).\label{eq:sup_kw}
\end{eqnarray}
We can see that in the hybridized bands, there are actually \textit{two} pairs of degenerate bands sitting at the 
momenta ${\bf k}_W$.
Near ${\bf k}_W$,
the band dispersions 
can be linearized.

Due to the constraints in the strong-coupling regime, $\mathcal{H}^{s}_a$ must be solved self-consistently with the saddle-point equations,
\begin{equation}
\begin{array}{lr}
	\frac{1}{2 N_u} \sum_{{\bf k},a=\pm} \left\la \breve{\xi}^\dagger_{{\bf k}a} \breve{\xi}_{{\bf k}a}\right\ra +r^2 = 1, &~~~~~
	\frac{V}{4N_u} \sum_{{\bf k},a=\pm} \left[  \left\la \breve{\psi}^\dagger_{{\bf k}a} \breve{\xi}_{{\bf k}a}\right\ra + \Hc\right] + r \ell = 0,
\end{array}
\end{equation}
where $N_u$ is the number of the unit cell. Here we can tune the chemical potential to be $\mu = - V_s^2/E_s$, with $E_s>0$. This fixes the lower Weyl node to the Fermi energy, at $1/4$-filling,
as shown in 
Fig.~2 of the main text. For the illustration in the main texts, we use the same bare coupling parameters 
to be $(t,\lambda, E_d, V) =(1, 0.5, 1, -6, 6.6)$, and solved self-consistently for $r \simeq 0.259$ and $\ell \simeq 6.334$, 
with error $\epsilon\leq\mathcal{O}(10^{-5})$ on a $64 \times 64 \times 64$ unit cell diamond lattice. 

%%%%%%%%%%%%%%%%%
\section*{Berry curvature}
The Berry curvature field $\vec{\Omega}({\bf k})$ is akin to a ficticious magnetic field in momentum space; analogously, the Weyl nodes manifest as monopole sources and sinks of Berry curvature~\cite{Volovik_JETP, Bernevig13}.
The field is a way of representing the tensor components since it is a $3\times3$ 
antisymmetric tensor in three dimensions, $\vec{\Omega}({\bf k})=(\Omega_{yz}({\bf k}),\Omega_{zx}({\bf k}),\Omega_{xy}({\bf k}))$.
The components are given by the gauge invariant equation~\cite{Volovik_JETP, Bernevig13},
\begin{eqnarray}
\Omega_{ab}({\bf k})=\sum_{n\neq n'}\mathcal{I}m\frac{ \la n {\bf k}|\partial_{c,k_a} H^{s}_{\bf k}|n' {\bf k}\ra\la n' {\bf k}|\partial_{c,k_b} H^{s}_{\bf k}|n {\bf k}\ra }{(\mathcal{E}_n - \mathcal{E}_{n'})^2},
\end{eqnarray}
where $H^s_{\bf k}$ is the $8\times8$ Bloch matrix in the strong coupling regime in physical spin space, $\partial_{c,k_a}$ is the derivative with respect to \textit{only} the conduction electrons corresponding to the velocity of the charge carriers. 
$\mathcal{E}_n=\mathcal{E}^{(\tau,\alpha)}_\nu$ and $|n{\bf k}\ra$ are the Bloch eigenenergies and eigenstates of $H^{s}_{\bf k}$, with index $n$ specifying one of the eight bands, $n=(\tau,\alpha,\nu)=(\pm,\pm,\pm D)$.

In the main text, the Berry curvature of the heavy Weyl fermions in the strong coupling regime were shown; specifically we plot the field's unit length 2D projection onto the $k_x$-$k_y$ plane, 
\begin{eqnarray}
	\hat{\Omega}(k_x,k_y,2\pi)=\frac{1}{|\vec{\Omega}(k_x,k_y,2\pi)|}\begin{pmatrix}\Omega_{yz}(k_x,k_y,2\pi),~\Omega_{zx}(k_x,k_y,2\pi)\end{pmatrix}.
\end{eqnarray}
%%%%%%%%%%%%%%%%

\section*{Surface states}
Following the approach in Ref.~\cite{Ojanen13}, we also seek surface states in the $|-D\ra$ sector near the Weyl nodes. The nodes are on the square faces of the fcc Brillouin zone boundary, along the lines connecting high symmetry points $X$ and $W$. We find that the Hamiltonian matrix of $\mathcal{H}^{s}_-$, 
%Eq.~
\eqref{Seq:scpm} can be expressed
\begin{eqnarray} 
\nonumber h_{-} ({\bf k})&&=(\kappa^0+\kappa^z) \otimes \tfrac{1}{2} \left[ u_1({\bf k})\tau_x + u_2({\bf k})\tau_y + (m -\lambda D({\bf k}))\tau_z)\right]\\
&&+\left[(E_s -\mu)\kappa^0-(E_s + \mu )\kappa^z+V_s \kappa^x \right]\otimes\tau_0
\end{eqnarray}
where $\kappa^i$ are Pauli matrices acting on the $\begin{pmatrix} \breve{\psi}_{{\bf k}-},\breve{\xi}_{{\bf k}-}\end{pmatrix}$ basis. 
We linearize the Hamiltonian matrix near ${\bf q} = \begin{pmatrix} k_x, k_y, 2\pi \end{pmatrix}$ in $\tilde{q}_z = k_z - 2\pi$ around $\tilde{q}_z = 0$. 
Defining $u'_{1/2} \equiv \partial_{k_z} u_{1/2}({\bf k})|_{k_z = 2\pi}$, $u' \equiv \sqrt{(u'_1)^2 + (u'_2)^2}$, and $u \equiv \sqrt{u_1^2 + u_2^2}$, we obtain
\begin{eqnarray}
	\nonumber h_{-}({\bf q}) &&= (\kappa^0+\kappa^z) \otimes \tfrac{1}{2} [  \tilde{q}_z u'({\bf q})\tau_x - u({\bf q})sgn(k_xk_y)\tau_y + (m -\lambda D({\bf q}))\tau_z ]\\	
&&+\left[(E_s -\mu)\kappa^0 -(E_s + \mu )\kappa^z +V_s \kappa^x \right]\otimes\tau_0.
\end{eqnarray}
Making the real-space replacement $\tilde{q}_z \rightarrow -i \partial_z$, we can enforce the boundary condition by assigning the value of the staggered mass to be $m = m_+ >4|\lambda|$ outside for $z>0$ (trivially insulating vacuum), and $m= m_- <4|\lambda|$ for $z <0$, such that the bulk is in a stable Weyl semimetal phase.

Generalizing the wavefunction suggested in Ref.~\cite{Ojanen13}, we find the surface eigenstates in the plane of the four Weyl nodes around the $X$ point to be
\begin{eqnarray}
\psi_s (k_x, k_y, z) &&= A_s e^{\mp \frac{z}{\xi_\pm}} | \kappa \ra\otimes| \tau_y = +1\ra \otimes | - D\ra,~~~~~
\end{eqnarray}
with
\begin{eqnarray}
| \kappa \ra &&= B_s 
\begin{pmatrix} 
1 \\
- \sqrt{1 - \left( \frac{\beta^s_{\bf k}}{V_s}\right)^2} - \frac{\beta^s_{\bf k}}{V_s}
\end{pmatrix},\\
\beta^s_{\bf k} &&\equiv  -\frac{1}{2} \left[u({\bf q}) sgn(k_x k_y) + E_s + \mu \right],
\end{eqnarray}
where $A_s$, $B_s$ are normalization constants. We identify $\xi_\pm = \pm u'({\bf q}) /(m_{\pm}-\lambda D({\bf q}))$ as the penetration depth of the surface wavefunctions. Inside the boundary, the divergence of $\xi_-$ when $D({\bf q}) = m_-$ indicates that the surface states merge with the bulk states, becoming indistinguishable~\cite{Ojanen13}.

%%%%%%%%%%%%%%%%%%%%%%%%%%%%%%%%%%%%%%%
\section*{
Inversion-symmetry breaking and time-reversal-symmetry breaking cases}
Here we consider $H_c$ in the presence of a static magnetic field, as an illustration of the effect of a time-reversal symmetry breaking on the Weyl state. Consider 
the conduction electron Hamiltonian,
\begin{eqnarray}
	H_c &&= \sum_{\la i j \ra, \sigma } \left( t_{ij} c_{i \sigma}^\dagger c_{j \sigma}+ \Hc \right) + i \lambda \sum_{\la\la i j \ra\ra} \left[ c^\dagger_{i \sigma} \left( {\bm{\sigma}}\cdot {\bf e}_{ij} \right) c_{j\sigma} - \Hc\right]\\
	&&+ m \sum_{i,\sigma} (-1)^i c^\dagger_{i\sigma} c_{i\sigma}
	+ \sum_j \bm{M} \cdot \left( c^\dagger_{j\sigma}  \bm{\sigma} c_{j\sigma} \right)
\end{eqnarray}
where ${\bf e}_{ij} = \frac{{\bf e}_i \times {\bf e}_j}{| {\bf e}_i \times {\bf e}_j |}$ are determined by the two bond vectors connecting second-nearest-neighbors, and $\bm{\sigma} = (\sigma_x,\sigma_y,\sigma_z)$ are the Pauli matrices acting on spin space, and the last term is the time reversal symmetry breaking (TRB) term, with $\bm{M}$ being the local moment and $c^\dagger_{j\sigma} \bm{\sigma} c_{j \sigma}$ being the conduction electron spin.

\begin{figure}[t]
   \centering
    \subfigure[]{\label{Fig:zincblende}\includegraphics[width= 2 in]{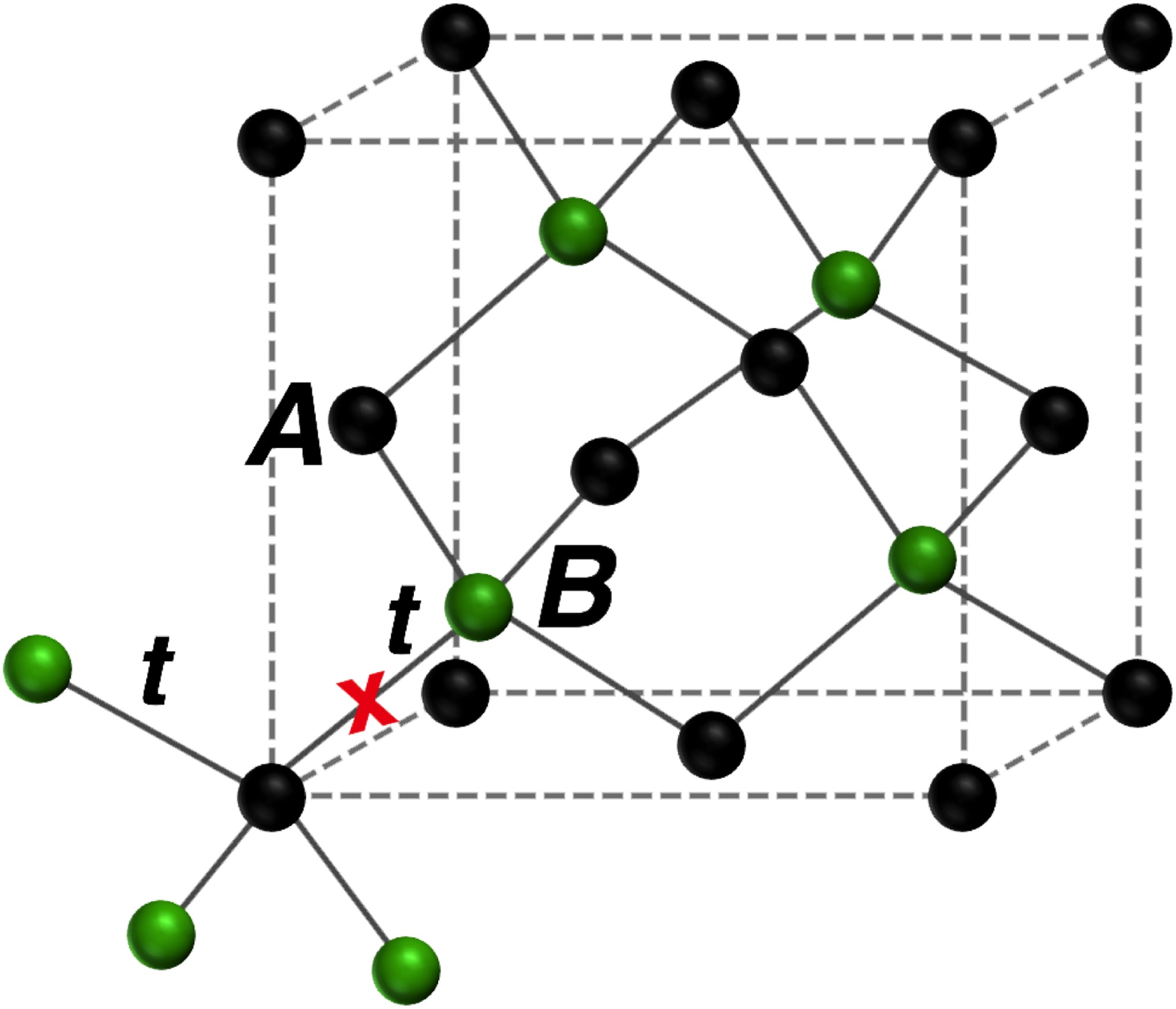}}
    \subfigure[]{\label{Fig:inv_dia}\includegraphics[width= 1.5 in]{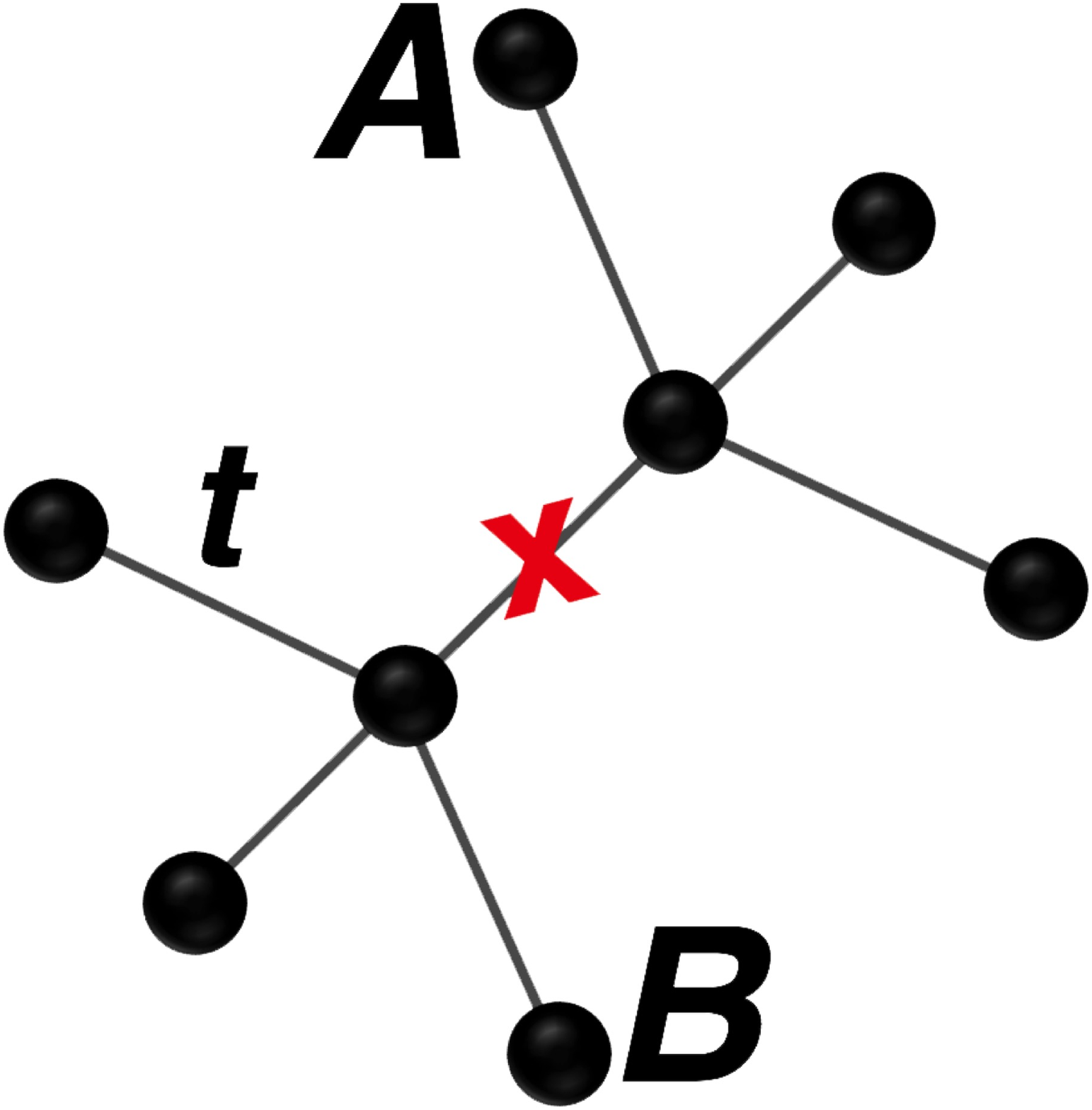}}
    \subfigure[]{\label{Fig:inv_zincb}\includegraphics[width= 1.5 in]{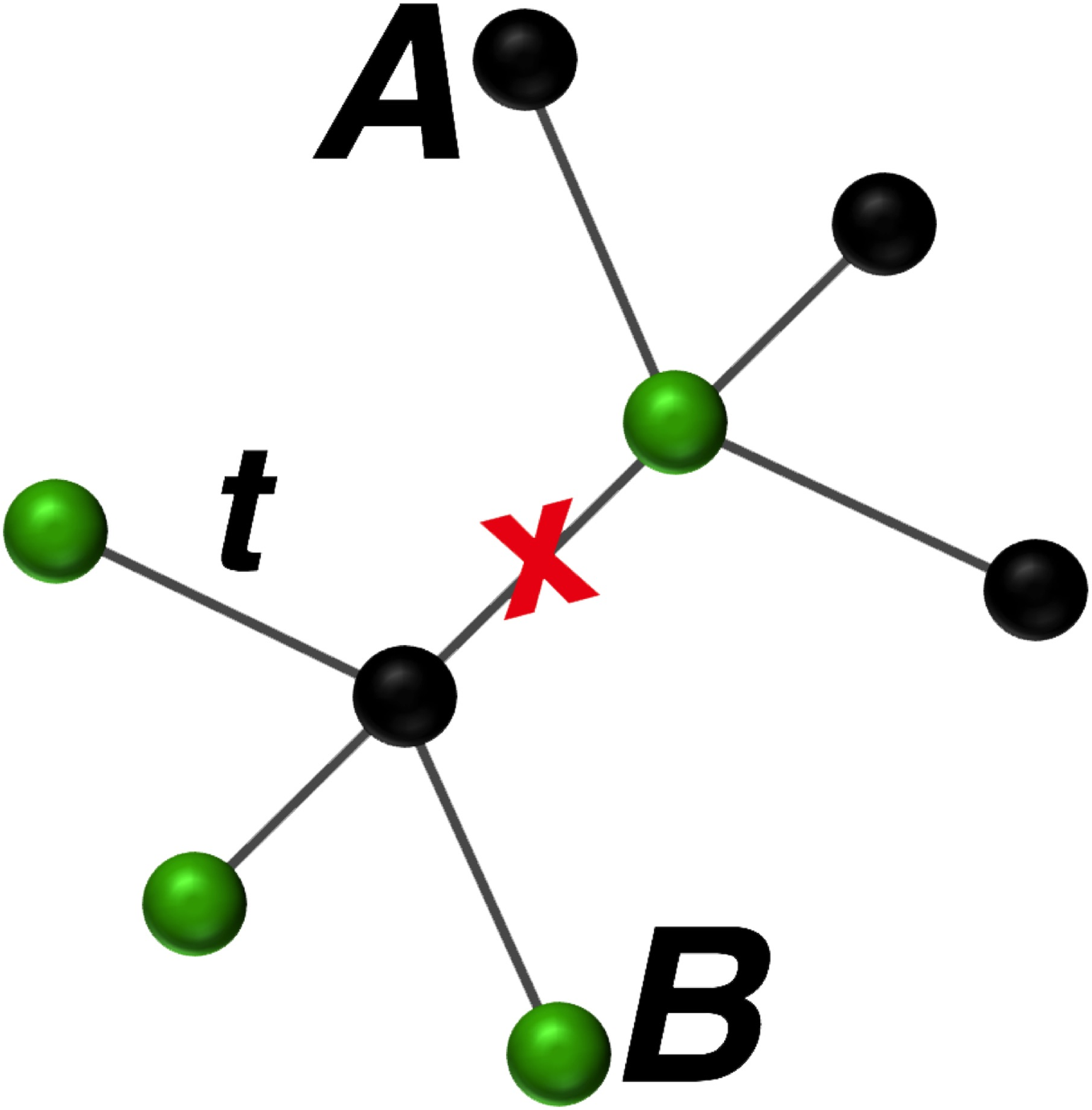}}
    \caption{(a) Zincblende lattice: diamond lattice with $\pm m$ differentiating $A,~B$ sublattice; 
    (b) diamond lattice unit cell, and 
(c) zincblende unit cell.
Translating an $A$ atom across the inversion point marked ``X''
exchanges it with the B site, breaking inversion symmetry in (a) and (c), but preserving it in (b) since the two sites are equivalent.}
\label{Fig:lattice_zincblende}
\end{figure}

The noncentrosymmetric diamond lattice (the ``zincblende'' lattice) is presented in Fig.~S\ref{Fig:zincblende}.
Although the diamond and zincblende lattices are structurally the same, the different on-site potential $m$ reduces the $O_h$ cubic point group symmetry of the diamond lattice to tetragonal $T_d$ symmetry in the zincblende.
The simplest way to visualize the inversion symmetry-breaking introduced by the $m$ term is to compare the interlocked 
sublattice unit cells under the inversion operations in Fig.~S\ref{Fig:inv_dia}. If one reflects the position of any site 
across the indicated 
inversion center `X',
the upper four sites neatly exchange 
positions with the lower four.
Conversely, if the on-site potential differentiates the sublattices via $m$, an inversion operation exchanges the distinct sublattice sites, 
as seen in Fig.~S\ref{Fig:inv_zincb}, so the inversion symmetry is broken.

\begin{figure}[t]
   \centering
    \subfigure[]{\label{Fig:supp_iso_dirac}\includegraphics[width= 2 in]{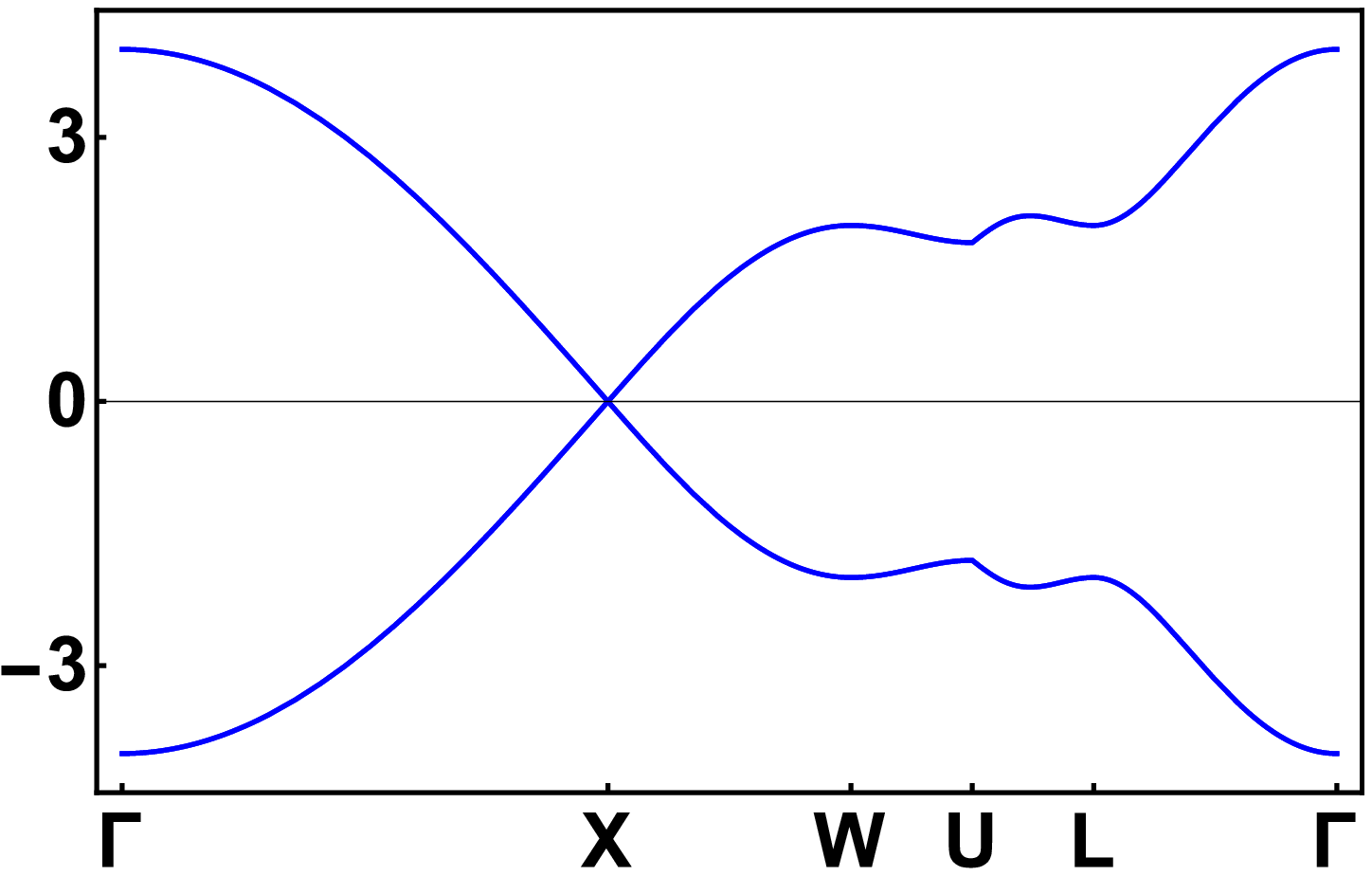}}
    \subfigure[]{\label{Fig:supp_iso_xgamma3}\includegraphics[width= 2 in]{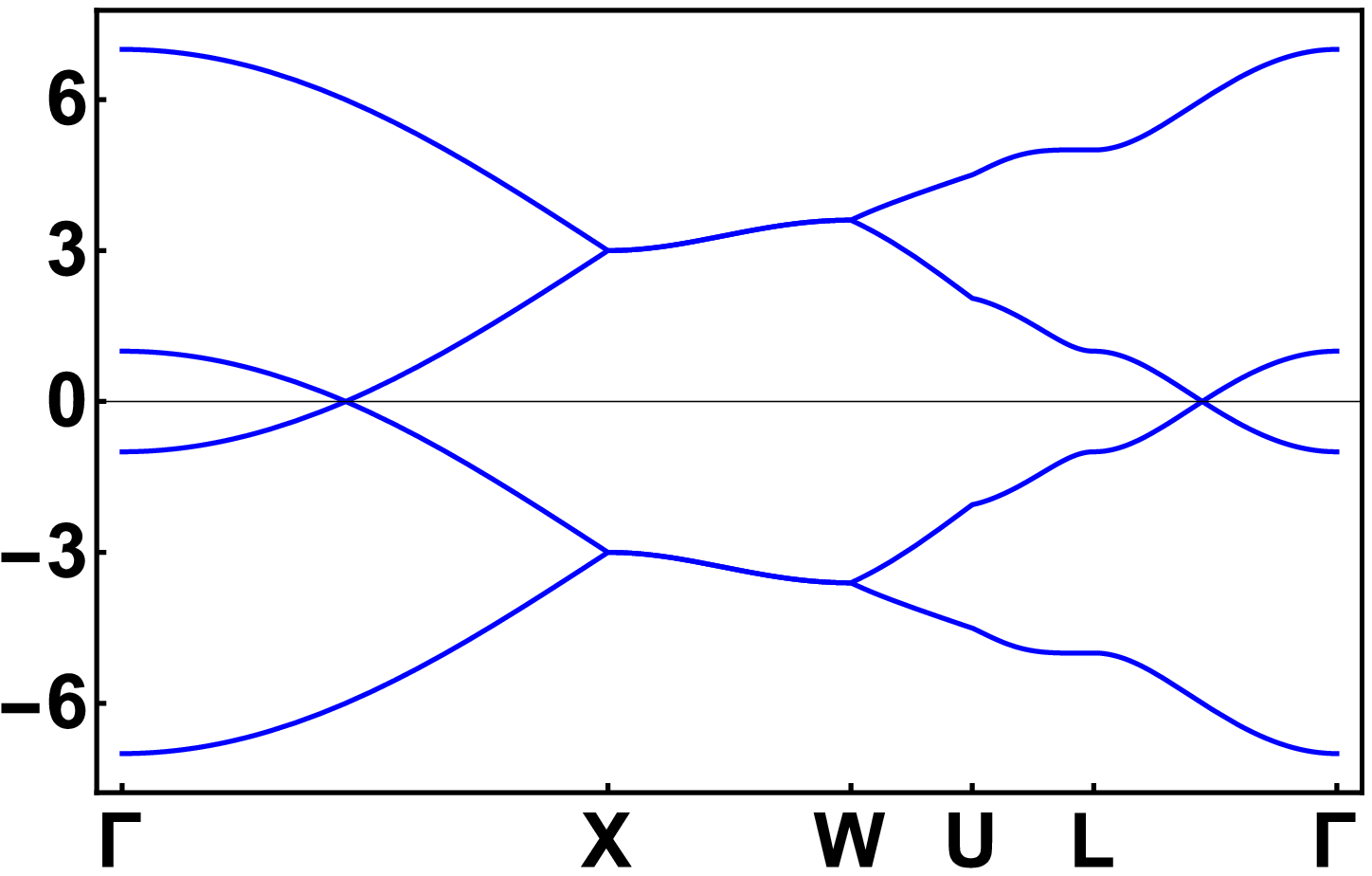}}
    \caption{
    Band structure along the high symmetry points of the fcc Brillouin zone in the presence of time-reversal symmetry breaking (preserving inversion symmetry), originated from a local moment field $M_z$ coupled to the conduction electron spin, using $\lambda=\frac{1}{2}$.
    (a) $M_z = 0$: Dirac semimetal; 
    (b) $M_z =3$: Weyl nodes along different lines are all moved toward $\Gamma$ point. }
\label{Fig:TRB_iso}
\end{figure}

Here we show an example of a Weyl semimetal phase in the broken time reversal symmetry $(\bm{M} \neq0)$, 
in the case of 
the diamond lattice $(m=0)$. For simplicity, below we choose $\bm{M} = M_z \hat{z}$ and the second term becomes a Zeeman-like term. Following the same procedure as before, we introduce the basis in momentum space $\Psi^T_{\bf k} = \begin{pmatrix} c_{{\bf k}\uparrow, A} & c_{{\bf k} \uparrow, B} & c_{{\bf k}\downarrow, A} & c_{{\bf k}_\downarrow, B} \end{pmatrix}$. The Hamiltonian of the conduction electron becomes $H_{c,M}^{TRB} = \Psi^\dagger_{\bf k} \cdot h^{TRB}_c \cdot \Psi_{\bf k}$, with 
\begin{eqnarray}
h^{TRB}_c =  \sigma_0 \left[ u_1({\bf k}) \tau_x + u_2({\bf k}) \tau_y \right] + M_z \sigma_z \tau_0 + \lambda\left[ {\bf D}({\bf k}) \cdot \bm{\sigma}\right] \tau_z,
\end{eqnarray}
 where $u_{1/2}$ and vector ${\bf D}$ are defined in the main text. 
 A 3D Dirac semimetal is realized with Dirac points located at $X$ points in the absence of the TRB term, 
 as illustrated in Fig.~S\ref{Fig:supp_iso_dirac}.  

 %%%%%
%%%%%%
Upon increasing the TRB term, first we observe that Weyl nodes appear along $X-\Gamma$ lines and along $X-W$ lines on the 3D BZ boundary parallel to $\hat{k}_z$-axis, and increasing to $M_z=3$ moves both Weyl nodes toward the $\Gamma$ point, illustrated in Fig.~S\ref{Fig:supp_iso_xgamma3}. 

Next, we analyze the critical value of $M_z$ signaling the phase transition that separates the aforementioned TRB-Weyl semimetals from the topologically trivial insulator phase. 
Focusing on one of the $X-\Gamma$ lines that we suspect to harbor Weyl nodes, we assume its position is at ${\bf k}_W = {\bf k}_X - \delta{\bf k}$, where ${\bf k}_X$ is the $X$-point momentum, and $\delta{\bf k}=(\delta k_x,0,0)$ with $|\delta k_x| >0$. 
We can then straightforwardly find that all the components of ${\bf D}({\bf k})$ along that line vanish. 
The Hamiltonian matrix along the line can be simplified to be
\begin{eqnarray}
\nonumber h_c^{TRB}|_{X \Gamma} =  2t \sigma_0 \left[ \left( 1-\cos\left(\frac{\delta k_x}{2}\right)\right)\tau_x + \sin\left( \frac{\delta k_x}{2}\right) \tau_y \right]  +M_z \sigma_z \tau_0  ,
\end{eqnarray}
which leads to the eigenvalues
\begin{eqnarray}
E^{TRB,\sigma}_c = \sigma M_z \pm 2\sqrt{2 \left[ 1 - \cos \left( \frac{\delta k_x}{2}\right) \right]},
\end{eqnarray}
where $\sigma=\pm$ for spin=$\up,\dn$. For $M_z >0$, we can see that gaplessness can only occur when $M_z = 2 \sqrt{2 \left[ 1 - \cos \frac{\delta k_x}{2}\right]}$, which leads to the condition
\begin{eqnarray}
\cos\left(\frac{\delta k_x}{2} \right) = 1 - \frac{M_z^2}{8}.
\end{eqnarray}
Therefore, we can see that the condition can be satisfied for $0<M_z \leq M_z^c = 4$, after which the absolute value of the right hand side becomes greater than $1$ and the condition can no longer be held. The critical value of $M_z^c = 4$ is fully consistent with the numerical analysis 
 illustrated in Fig.~\ref{Fig:TRB_iso}.

%%%%%%%%%%%%%%%%%%%%%%%%%%%%%%%%%%%%%%%
\section*{Specific heat from a Weyl node}
The specific heat is calculated as
\begin{eqnarray}\label{eq:spec-1}
c_v&&=\left(\frac{\partial u}{\partial T} \right)_V = \frac{\partial}{\partial T}\int_{BZ}\frac{d^3{\bf k}}{(2\pi)^3}\varepsilon_{\bf k}f(\varepsilon_{\bf k}),
\end{eqnarray}
where $\varepsilon_{\bf k}$ is the energy dispersion, $u$ is the energy density, and the integral is over the first Brillouin zone.
Here the occupation distribution function $f(\varepsilon_{\bf k})$ is the Fermi function.
We focus on the linear dispersion regime where we can approximate $\varepsilon_{\bf k}=\hbar v^* k$. 
We will take the renormalized Fermi velocity $v^*$ for its asymptotic low-temperature value. Analyzing the 
 temperature dependence of the condensate amplitude in our saddle-point analysis will only cause subleading corrections to the temperature
 dependence of the specific heat.

The result for the specific heat 
per unit volume is
\begin{eqnarray}\label{eq:spec-5}
c_v =\frac{7\pi^2}{30}k_B\left(\frac{k_BT}{\hbar v^*}\right)^3.
\end{eqnarray}
This shows 
 that the $T^3$ contribution to the specific heat becomes large when $v^*$ is small, as in the case of heavy fermion systems.
 
In the above calculation, we have assumed that the leading term of the specific heat at low temperatures 
is independent of the quasiparticle weight  or the residual interactions of the nodal excitations. 
This is because the entropy counts the number of the degrees of freedom that are thermally excited within an energy range of about $k_BT$. To put this argument on a more concrete footing, we turn to an alternative calculation.

%%%%%%%%%%%%%%%%%%%%%%%%%%%%%%%%%%%%%%%%
\section*{Fermi liquid 
approach  
to the specific heat of a Weyl fermion}
Adopting 
the Fermi-liquid approach for calculating the entropy, as outlined in Ref.~\cite{Book_AGD} 
(chapter 4, section 19), 
we  express the specific heat per unit volume from a Weyl fermion as
\begin{eqnarray}
c_v =  \int \frac{d^3{\bf k}}{(2\pi)^2} \frac{1}{2\pi i T} \int_{-\infty}^{\infty} \varepsilon \left[ - \frac{\partial f(\varepsilon)}{\partial \varepsilon}\right] \left[ \ln G_R({\bf k}, \varepsilon) - \ln G_A({\bf k}, \varepsilon) \right] d\varepsilon,
\end{eqnarray}
where $T$ is the temperature, $k_B$ is the Boltzmann constant, and $f(\varepsilon)$ is the Fermi distribution function. In addition, 
$G_R({\bf k},\varepsilon) = Z/(\varepsilon - \xi_{\bf k} + i 0^+)$, is the retarded Green function of Fermi liquid quasi-particles, 
with $Z$ being the quasi-particle weight, and $\xi_{\bf k} = \hbar v^* |{\bf k}| \equiv \hbar v^* k$ for $\xi_{\bf k} >0$ ($\xi_{\bf k} 
= -\hbar v^* k$ for $\xi_{\bf k} <0$), \textit{i.e.}, a Weyl fermion dispersion.
Likewise, the advanced Green function is $G_A= G_R^*$, and we set the chemical potential $\mu=0$, sitting exactly at the nodal point. 
We use the Sommerfeld expansion 
\begin{eqnarray}
\int_{-\infty}^\infty d\varepsilon u(\varepsilon) \left[ \frac{\partial f(\varepsilon)}{\partial \varepsilon}\right] \simeq - u(0) 
- \frac{\pi^2}{6} (k_B T)^2 \left( \frac{\partial^2 u(\varepsilon)}{\partial \varepsilon^2}\right)_{\varepsilon =0} 
- \frac{7\pi^4}{360} (k_B T)^4 \left( \frac{\partial^4 u(\varepsilon)}{\partial \varepsilon^4} \right)_{\varepsilon = 0} + \cdots.
\end{eqnarray}
Taking derivatives and setting $\varepsilon = 0$, we obtain 
\begin{eqnarray}
c_v =  \int \frac{d^3{\bf k}}{(2\pi)^3} k_B \left[ \frac{\pi}{3} k_B T~ Im\left( G_R^{-1} \partial_\varepsilon G_R \right)_{\varepsilon = 0} + \frac{7\pi^4}{90} (k_BT)^3  Im \left( G_R^{-1} \partial_\varepsilon^3 G_R + 2 G_R^{-3} \left( \partial_\varepsilon G_R\right)^3 - 3 G_R^{-2} \partial_\varepsilon G_R \partial_\epsilon^2 G_R\right)_{\varepsilon =0} \right],
\end{eqnarray}
where $Im \left( A\right)$ means the imaginary part of $A$. For a Weyl fermion, we find that $Im\left[G_R^{-1} \partial_\varepsilon G_R \right] = \pi \delta(\varepsilon - \xi_{\bf k})$, where $\delta(x)$ is the Dirac delta function;  the identity
\begin{eqnarray}
\frac{1}{x + i 0^+} = \mathcal{P}_v \left( \frac{1}{x} \right) - i \pi \delta(x)
\end{eqnarray}
has been used, 
with $\mathcal{P}_v$ 
denoting principal value. For the leading linear-$T$ term, the integral involves
\begin{eqnarray}
\int \frac{4\pi k^2 dk}{(2\pi)^3}  \delta(\xi_{\bf k}) =0,
\end{eqnarray}
and therefore the leading linear-in-$T$ term vanishes. For the second term above, we find that
\begin{eqnarray}
\nonumber &&  G_R^{-1} \partial^3_\varepsilon G_R + 2 G^{-3}_R \left( \partial_\varepsilon G_R\right)^3 - 3 G^{-2}_R \partial_\varepsilon G_R \partial^2_\varepsilon G_R = -2 \left[ \mathcal{P}_v \left(\frac{1}{\varepsilon - \xi_{\bf k}}\right) - i \pi \delta(\varepsilon- \xi_{\bf k})\right]^3  \\
&& = -2 \left[ \mathcal{P}_v\left( \frac{1}{\varepsilon- \xi_{\bf k}} \right)^3 + 3 \mathcal{P}_v\left( \frac{1}{\varepsilon- \xi_{\bf k}} \right)^2 (-i \pi ) \delta(\varepsilon - \xi_{\bf k}) + 3 \mathcal{P}_v \left( \frac{1}{\varepsilon - \xi_{\bf k}}\right) (-i \pi)^2 \delta^2 (\varepsilon - \xi_{\bf k}) + \left(-i \pi \delta(\varepsilon - \xi_{\bf k})\right)^3\right]. \label{supp:eq_FL}
\end{eqnarray}  
Focusing on the imaginary terms, we find that only the second term in~\eqref{supp:eq_FL} contributes to the results. The last term vanishes because the integral of cubic delta function is zero. Therefore, we find that 
\begin{eqnarray}
Im \left( G_R^{-1} \partial_\varepsilon^3 G_R + 2 G_R^{-3} \left( \partial_\varepsilon G_R\right)^3 - 3 G_R^{-2} \partial_\varepsilon G_R \partial_\epsilon^2 G_R\right)_{\varepsilon =0} = 6\pi \mathcal{P}_v \left( \frac{1}{\xi_{\bf k}}\right)^2 \delta ( \xi_{\bf k}).
\end{eqnarray}
Combining
all these,
we
 obtain the specific heat per unit volume as
\begin{eqnarray}
c_v &\simeq & \frac{7 \pi ^4}{15} k_B (k_B T)^3 \int \frac{d^3 {\bf k}}{(2\pi)^3} \mathcal{P}_v \left( \frac{1}{\xi_{\bf k}}\right)^2 \delta (\xi_{\bf k})\\
& = & \frac{7 \pi^4}{90}k_B (k_B T)^3  \frac{1}{2} \int_{-\infty}^\infty  \frac{ 4\pi k^2 dk}{(2\pi)^3}  \frac{2}{ (\hbar v^* k)^2} \delta( \hbar v^* k) \label{supp:eq_cv}\\
& = & \frac{7 \pi^2}{30} k_B \left( \frac{ k_B T}{\hbar v^*}\right)^3 ,
\end{eqnarray}
\textit{i.e.}, the same expression 
as in Eq.~(\ref{eq:spec-5}).
It is worth noting that i) 
the principal value evaluation is not necessary since the $k^2$ in the denominator cancel the $k^2$ from the numerator;
ii) the factor $1/2$ in front of the integral in~\eqref{supp:eq_cv} is due to the extension of integration range
from $-\infty$ to $+\infty$ while recognizing that 
the function is even; and iii) the factor of 2 in the numerator of $2/(\hbar v^* k)^2$ inside the integral in~\eqref{supp:eq_cv} is originated
 from the fact that, for each momentum $k$, there are two contribution from $\xi_{\bf k} = \pm \hbar v^* k$ 
 in the Weyl fermion band distribution. 

In summary, the
Femi liquid analysis here 
demonstrates that, even when the Fermi-liquid effects (the quasiparticle weight and residual interactions of the nodal excitations) are 
explicitly taken
into account, the leading renormalization effect to the specific heat is still for the $T^3$ coefficient and 
has the form $(1/v^*)^3$.

\bibliography{biblio5KWSM}